%% file: ms2.tex
\documentclass[linenumbers]{aastex}
\usepackage{emulateapj5}
\newcommand{\inst}[1]{\altaffilmark{#1}}
\newcommand{\tablefoot}{\tablecomments}

\usepackage{natbib}
\bibpunct{(}{)}{;}{a}{}{,} 
\usepackage{amssymb}
\usepackage{url}
\usepackage{hyperref}
\usepackage{graphicx}
\usepackage{longtable}
\usepackage[usenames,dvipsnames]{color}
\usepackage{booktabs}
\usepackage{lscape}

\newcommand{\submm}{submillimeter}

\newcommand{\msun}{\mbox{$M_\odot$}}
\newcommand{\mstar}{\mbox{$M_*$}}
\newcommand{\mdust}{\mbox{$M_{\rm dust}$}}
\newcommand{\mgas}{\mbox{$M_{\rm gas}$}}

\newcommand{\msunyr}{\mbox{\msun\ yr$^{-1}$}}
\newcommand{\hi}{\ion{H}{1}}
\newcommand{\mhi}{\mbox{$M_{\rm H{\sc I}}$}}

\newcommand{\oiii}{[\ion{O}{3}]}
\newcommand{\lp}{\mbox{$L'$}}
\newcommand{\kms}{\mbox{km\,s$^{-1}$}}

\newcommand{\htwo}{\mbox{H$_2$}}
\newcommand{\mhtwo}{\mbox{$M_{\rm H_2}$}}
\newcommand{\fhtwo}{\mbox{$f_{\rm H_2}$}}

\urldef{\hyperleda}\url{http://leda.univ-lyon1.fr/ledacat.cgi?NGC%203278}

\newcommand{\ud}{\mathrm{d}}  

\defcitealias{michalowski19etg}{M19}
\defcitealias{rowlands12}{R12}

\begin{document}

 \title{The fate of the interstellar medium in early-type galaxies. III. The mechanism of interstellar medium removal and the quenching of star formation
 }
 

\shorttitle{ISM in ETGs III. mechanism of  removal and quenching}
\shortauthors{Micha{\l}owski et al.}

\author{Micha{\l}~J.~Micha{\l}owski\inst{\ref{inst:poz},\ref{inst:cal},\ref{fulb},\ref{inst:roe}}%
, 
C.~Gall\inst{\ref{inst:dark}}%
, 
J.~Hjorth\inst{\ref{inst:dark}}%
, 
D.~T.~Frayer\inst{\ref{inst:nrao}}%
, 
A.-L.~Tsai\inst{\ref{inst:poz}}%
, 
K.~Rowlands\inst{\ref{inst:stsi},\ref{inst:jhu}}%
, 
T.~T.~Takeuchi\inst{\ref{inst:nagoya},\ref{inst:ism}}%
, 
A.~Le\'{s}niewska\inst{\ref{inst:poz}}%
, 
D.~Behrendt\inst{\ref{inst:roe}}%
, 
N.~Bourne\inst{\ref{inst:roe}}%
, 
D.~H.~Hughes\inst{\ref{inst:inaoe}}%
, 
M.~P.~Koprowski\inst{\ref{inst:umk}}%
, 
J.~Nadolny\inst{\ref{inst:poz}}%
, 
O.~Ryzhov\inst{\ref{inst:poz}}%
, 
M.~Solar\inst{\ref{inst:poz}}%
, 
E.~Spring\inst{\ref{inst:roe}, \ref{inst:ams}}%
, 
J.~Zavala\inst{\ref{inst:naoj}}%
, 
P.~Bartczak\inst{\ref{inst:poz},\ref{inst:al}}	
}

\altaffiltext{1}
{Astronomical Observatory Institute, Faculty of Physics, Adam Mickiewicz University, ul.~S{\l}oneczna 36, 60-286 Pozna{\'n}, Poland, {\tt mj.michalowski@gmail.com}\label{inst:poz}}
\altaffiltext{2}
{TAPIR, Mailcode 350-17, California Institute of Technology, Pasadena, CA 91125, USA  \label{inst:cal}}
\altaffiltext{3}
{Fulbright Senior Award Fellow \label{fulb}}
\altaffiltext{4}
{SUPA (Scottish Universities Physics Alliance), Institute for Astronomy, University of Edinburgh, Royal Observatory, Blackford Hill, Edinburgh, EH9 3HJ, UK  
\label{inst:roe}}
\altaffiltext{5}
{DARK, Niels Bohr Institute, University of Copenhagen, Jagtvej 128, DK-2200 Copenhagen N, Denmark  
\label{inst:dark}}
\altaffiltext{6}
{National Radio Astronomy Observatory, P.O. Box 2, Green Bank, WV 24944, USA \label{inst:nrao}}
\altaffiltext{7}
{AURA for ESA, Space Telescope Science Institute, 3700 San Martin Drive, Baltimore, MD, USA \label{inst:stsi}}
\altaffiltext{8}
{Department of Physics and Astronomy, Johns Hopkins University, Baltimore, MD 21218, USA \label{inst:jhu}}
\altaffiltext{9}
{Division of Particle and Astrophysical Science, Nagoya University, Furo-Cho, Chikusa-ku, Nagoya 464-8602, Japan \label{inst:nagoya}}
\altaffiltext{10}
{The Research Center for Statistical Machine Learning, the Institute of Statistical Mathematics, 10--3 Midori-cho, Tachikawa, Tokyo 190-8562, Japan \label{inst:ism}}
\altaffiltext{11}
{Instituto Nacional de Astrof\'{\i}sica, \'Optica y Electr\'onica (INAOE), Aptdo. Postal 51 y 216, 72000 Puebla, Pue., Mexico\label{inst:inaoe}}
\altaffiltext{12}
{Institute of Astronomy, Faculty of Physics, Astronomy and Informatics, Nicolaus Copernicus University, Grudzi\c{a}dzka 5, 87-100 Toru\'{n}, Poland
\label{inst:umk}}
\altaffiltext{13}
{Anton Pannekoek Institute, University of Amsterdam, Science Park 904, NL-1098 XH Amsterdam, the Netherlands\label{inst:ams}}
\altaffiltext{14}
{National Astronomical Observatory of Japan, 2-21-1 Osawa, Mitaka, Tokyo 181-8588, Japan \label{inst:naoj}}
\altaffiltext{15}
{Instituto Universitario de F\'{i}sica Aplicada a las Ciencias y las Tecnologías (IUFACyT). Universidad de Alicante, Ctra. San Vicente del Raspeig s/n. 03690 San Vicente del Raspeig, Alicante, Spain
\label{inst:al}}

\begin{abstract}
{
Understanding how galaxies quench their star formation is crucial for studies of galaxy evolution. 
Quenching is related to a decrease of cold gas. 
In the first paper 
we showed that the dust removal timescale in early-type galaxies (ETGs) 
is about 2.5\,Gyr. Here we present carbon monoxide (CO) and 21 cm hydrogen ({\hi}) line observations of these galaxies and measure the timescale of removal of the cold interstellar medium (ISM). We find that all the cold ISM components (dust, molecular and atomic gas) decline at similar rates. This allows us to rule out a wide range of potential ISM removal mechanisms (including starburst-driven outflows, astration, a decline in the number of asymptotic giant branch stars), and artificial effects like stellar mass-age correlation, environmental influence, mergers, and selection bias, leaving ionization by evolved low-mass stars and ionization/outflows by supernovae Type Ia or active galactic nuclei as viable mechanisms. We also provide evidence for an internal origin of the detected ISM. Moreover, we find that the quenching of star formation in these galaxies cannot be 
explained by a reduction in the gas amount alone,
because the star formation rates (SFRs) decrease faster (on a timescale of about 1.8 Gyr) than the amount of cold gas. Furthermore, the star formation efficiency of the ETGs ($\mbox{SFE}\equiv\mbox{SFR}/\mhtwo$) is lower than that of star-forming galaxies, whereas their gas mass fractions ($\fhtwo\equiv \mhtwo/\mstar$) are normal. This may be explained by the stabilization of gas against fragmentation, for example due to morphological quenching, turbulence, or magnetic fields.
}
\end{abstract}



\section{Introduction}
\label{sec:intro}

In order to have a full picture of galaxy evolution, we need to understand how galaxies become passive, i.e., how they stop forming stars, the process called quenching.
Star formation ceases either when gas is removed from a galaxy or is made unable to form stars.
A galaxy can run out of cold gas  
when it is used for star formation \citep[astration;][]{schawinski14,peng15}. On the other hand, gas can be expelled or ionized by either supernovae (SNe; \citealt{dekel86,ceverino09,muratov15,hopkins18,li20}) or evolved low-mass stars \citep{binette94,conroy15,herpich18,hopkins18b}.
Active galactic nuclei (AGN) have been claimed to be responsible for heating and removing cold gas, and suppressing star formation in more massive galaxies \citep{dimatteo05,springel05b,fabian12,cheung16b,bluck16,bluck20,hopkins16,piotrowska22}. The bulge of a galaxy may also make the gas resilient against fragmentation, shutting down star formation  \citep[i.e.~morphological quenching;][]{martig09,martig13,bluck14,bluck20b,bitsakis19,lin19,gensior20}.
A similar effect can result from turbulence and magnetic fields \citep{padoan02,federrath12}.  Finally, gas may be expelled from a galaxy as a result of interactions with other galaxies and mergers \citep{mcgee11,bekki14,davies15,davies19,poggianti17,sazonova21}.

The existence of gas in many passive galaxies may contradict the interstellar medium (ISM) removal as the only mechanism of quenching. Indeed,
atomic and molecular gas and dust have been detected for a fraction of early-type galaxies \citep[ETGs;][]{davis11,rowlands12,davis16b,davis19,davis16,young11,smith12,alatalo13,alatalo15,diseregoalighieri13,ashley17b, ashley18,ashley19,zhang19,richtler20,magdis21,donevski23} and post-starburst galaxies \citep{french15,rowlands15,alatalo16,suess17,yesuf17,smercina18,smercina22,li19,yesuf20,bezanson22,otter22,wu23,zanella23}. 
Small amounts of dense gas, traced by the hydrogen cyanide (HCN) line, in post-starburst galaxies \citep{french18b,french23} may point at the inability of the gas reservoir to collapse and form stars. 
This may be due to morphological quenching.

The origin of the ISM in ETGs remains unclear.
One possibility is that it is of internal origin, e.g., the left-overs from past star formation or released by low-mass stars \citep{knapp92,rowlands12,michalowski19etg}. 
If gas is brought in from the outside, then the orientation of its rotation is expected to be random with respect to the kinematic axes of a galaxy. However,  analysing the kinematic misalignment of stellar and gas components in ETGs, \citet{davis16} found that the paucity of counter-rotating gas disks implies very short gas depletion rates and unrealistically high merger rates (in order to match the gas detection rate which would otherwise be low for short depletion times). Alternatively, a very long gas relaxation timescale must be invoked,  which is consistent with cosmological simulations \citep{vandevoort15}.
The alignment of the gas/dust disk and stellar component has also been used to argue against an external origin of the ISM in ETGs \citep{bassett17,sansom19,richtler20}. Moreover, \citet{griffith19} found that the stellar and gas metallicities of ETGs are similar, suggesting an internal origin of gas. 
Finally, \citet{babyk19} found that the molecular gas mass in ETGs is correlated with their hot gas mass, 
also suggesting an internal origin.
In a similar vein, 
cold gas in simulated ETGs comes from cooling from the hot halo \citep{lagos14b}.

The other possibility is an external origin of ISM. In this scenario the ISM is acquired by ETGs by mergers with gas-rich dwarf galaxies or gas inflows.
\citet{davis19b} found that only 7\% of ETGs have a lower gas metallicity than stellar metallicity (a clear signature of an external origin of gas), but given very short enrichment timescale (and hence short visibility of low-metallicity features), they estimated that for at least a third of ETGs the gas is of external origin.
Moreover, ETGs with dust lanes contain cold ISM, which has also been shown to be brought in by minor mergers  \citep{davis15}.
An external source of the ISM has been claimed for around half of ETGs, which have misaligned stellar and gas disks \citep{davis11,barreraballesteros14,barreraballesteros15,jin16,bryant19}.
This
has also been supported by other works \citep{young14,woodrum22,cao22,lee23}.
This topic has been investigated in simulations by \citet{lagos15} who found that a misalignment of the rotation axis of the gas and stellar components is mostly the consequence of cold gas flows.

This is the third paper in a series in which we analyze ISM removal from ETGs. 
We use the term ETG for galaxies which are morphologically classified as ellipticals, lenticulars (S0), or early-type spirals (Sa and SBa).
In \citet[][Paper I, \citetalias{michalowski19etg} hereafter]{michalowski19etg} we presented the decline of dust mass as a function of stellar age,  measurement of the dust removal timescale and the origin of dust in these galaxies.  
In \citet{lesniewska23} and Ryzhov et al. (in prep.) we present an expanded analysis of 2\,000 of these galaxies, allowing us to analyse which galaxy properties influence the dust decline.  In Nadolny et al. (in prep) we found similar galaxies in simulations, providing a physical insight into the mechanism of this process. 
The objectives of the present paper are: {\it i)} to determine the mechanism of the ISM decline in ETGs, and {\it ii)} to constrain the mechanism of quenching.

The paper is structured as follows. In Section~\ref{sec:sample} we present the ETG sample and in Section~\ref{sec:data} we describe our new CO and {\hi} data. 
Section~\ref{sec:mod} describes the numerical galaxy evolution model used to interpret the data.
We present the results in Section~\ref{sec:res}. 
We discuss the implication of the gas removal in ETGs on quenching of star formation in  Section~\ref{sec:removal}. In Section~\ref{sec:mechanism} we discuss possible mechanisms for this gas removal and in Section~\ref{sec:esource} we discuss the source of energy needed for this process.
We close with a summary of our results in Section~\ref{sec:conclusion}.
We use a cosmological model with $H_0=70$ km s$^{-1}$ Mpc$^{-1}$,  $\Omega_\Lambda=0.7$, and $\Omega_m=0.3$. We also assume a 
\citet{chabrier03} 
initial mass function (IMF), to which all star formation rates (SFRs) and stellar masses were converted (by dividing by 1.6) if given originally assuming the \citet{salpeter} IMF.
Errors are given as $1\sigma$.

\section{Sample}
\label{sec:sample}

As in \citetalias{michalowski19etg}, we use the sample of dusty ETGs from \citet{rowlands12}. This sample includes all galaxies with elliptical/lenticular morphology and red spirals from the {\it Herschel}  Astrophysical Terahertz Large Area Survey \citep[H-ATLAS;][]{hatlas} $\sim14\,\mbox{deg}^2$ Science Demonstration Field \citep{ibar10,pascale11,rigby11,smith11} that are detected at $250\,\mu$m. 
The specific selection criteria used by \citet{rowlands12} are:
\begin{enumerate}
\item Set within the H-ATLAS Science Demonstration Field. 
\item Matched to an optical Sloan Digital Sky Survey (SDSS) source with a spectroscopic redshift in
a range $0.01<z_{\rm spec}<0.32$  
within a $10''$ radius and with a match reliability greater than $0.8$ . 
\item {\it Herschel} $250\,\mu$m $>5\sigma$ detection. 
\item Visually classified by \citet{rowlands12} as early-type (elliptical or S0), or red spiral with near-ultraviolet (NUV) to $r$-band color of $\mbox{NUV} -r > 4.5$. The color selection has not been applied to ellipticals or lenticulars. 
\end{enumerate}

The sample consists of 61 galaxies, including 42 ellipticals or lenticulars, and 19 red spirals (mostly Sa or SBa). These galaxies will here be referred to as ETGs. Recently, \citet{zhou21} found that such red spirals have similar star formation histories as ellipticals, so they are treated collectively.  We used the galaxy properties derived by \citet{rowlands12} based on the spectral energy distribution modelling using the data from the Galaxy And Mass Assembly (GAMA) survey \citep{driver11,driver16,hill11,robotham10,baldry10}. 
In order to assess the AGN power, we also used the {\oiii} data from single fiber spectroscopy in the GAMA survey \citep{gordon17}\footnote{\url{www.gama-survey.org/dr3/data/cat/SpecLineSFR/}}.
Twelve of our galaxies are located within their respective star formation main sequence \citep{michalowski19etg}.

\section{Data}
\label{sec:data}

We observed the CO and {\hi} lines of 13 galaxies from the sample (nine ellipticals and four red spirals). We randomly selected them to have even coverage of the entire stellar age range. They also cover the entire stellar mass range from $10^{9.7}$ to $10^{11}\,\msun$.

\subsection{IRAM30m/EMIR: CO lines}

We performed observations with the IRAM 30-m telescope\footnote{Proposal no.~198-14, 62-15, and 174-15; PI: M.~Micha\l owski.}
using the Eight MIxer Receiver\footnote{\url{www.iram.es/IRAMES/mainWiki/EmirforAstronomers}} \citep[EMIR;][]{emir}. 
We implemented the wobbler switching mode (with the offset to the reference positions of 60\arcsec), which provides stable and flat baselines and optimizes the total observing time.
We centered one intermediate frequency (IF) at the frequency of the CO(1-0) line and the other at the frequency of the CO(2-1) line (the latter was not possible for all sources, see Table~\ref{tab:emir}).
We used the Fourier Transform Spectrometers 200 (FTS-200) backend providing 195\,kHz spectral resolution and 16\,GHz bandwidth in each linear polarization.  The observations were divided into 6-min scans, each consisting of 12 scans 30\,s long. The pointing was verified every 1--2 hr on a nearby quasar 0823+033. The observing log is presented in Table~\ref{tab:emir}. 
We reduced the data using the Continuum and Line Analysis Single Dish Software ({\sc Class}) package within the Grenoble Image and Line Data Analysis Software\footnote{\url{www.iram.fr/IRAMFR/GILDAS}} ({\sc Gildas}; \citealt{gildas}). Each spectrum for a given galaxy was calibrated, and corrected for baseline shape. Then all spectra were averaged.

The IRAM30m/EMIR spectra were binned to 30\,{\kms} channels. A Gaussian was fitted to the binned data, and in the case of detections the $2\sigma$ width of the Gaussian was adopted to integrate the line flux. In cases of non-detections, a $[-200, 200]\,\kms$ width was adopted. The error per spectral channel was calculated using the ranges $[-900, -400]\,\kms$ and $[400, 900]\,\kms$. Then the uncertainty of the flux estimation was calculated by a Monte Carlo simulation. We calculated the line luminosities based on eq.~3 in \citet{solomon97}. The molecular masses were calculated using $\alpha_{\rm CO}=5\,M_\odot\, (\mbox{K km s}^{-1} \mbox{ pc}^2)^{-1}$ (the conversion includes helium). The choice of the Galactic CO-to-{\htwo} conversion factor is justified by the fact that most ETGs have solar metallicity \citep{conroy14,davis19b}. The widths of the lines, integrated fluxes, luminosities and the resulting molecular gas masses are presented in Table~\ref{tab:emirres}. In addition to the results from the CO(1-0) line, the table shows those from the CO(2-1) line if such tuning was possible.
The CO spectra are shown in the left and middle columns of Fig.~\ref{fig:emirspec} in the Appendix. Out of 13 targets for CO(1-0) we detected nine, and out of nine targets for CO(2-1) we detected five.

\begin{table*}
\caption{Observing log for the IRAM30m/EMIR observations with integration times and 225\,GHz atmospheric opacity.}
\label{tab:emir}
\begin{center}
\begin{tabular}{llcc}
\hline\hline
Galaxy		& Obs. date 	& $t_{\rm int}$	& $\tau_{\rm 225\,GHz}$	\\
			&			&  (hr)		&		  \\
\hline
J085828.5+003814 & 2015 Jul 28 & 0.8  & 0.50--0.73 \\		
J085915.7+002329 & 2015 Jul 30, Oct 09, 2016 Mar 03 & 5.6   & 0.41--0.54,  0.19-0.35, 0.1--0.3  \\		
J085946.7$-$000020 & 2015 Jul 29, 30 & 2.4  & 0.70--0.82, 0.27-0.58  \\	
J090038.0+012810 & 2015 Oct 09, 10 & 	3.6 & 0.19--0.38, 0.48--0.58 \\ 		
J090234.3+012518 & 2015 Mar 22, 2016 Mar 04 & 3.4  & 0.27--0.32, 0.03-0.57  \\		
J090238.7+013253 & 2016 Mar 06 & 3.8 & 0.04-0.27 \\ 
J090312.4$-$004509 & 2015 Oct 10, Nov 18 & 4.8  & 0.60--0.73, 0.12-0.22  \\
J090352.0$-$005353 & 2015 Mar 22 & 0.8  & 0.25--0.38 \\		
J090551.5+010752 & 2015 Nov 19 & 4.4 & 0.16-0.29 \\			
J090718.9$-$005210 & 2015 Nov 22, 2016 Mar 01 & 4.0 & 0.13--0.35, 0.10--0.46 \\  
J090952.3$-$003019 & 2015 Nov 20, 22, 2016 Mar 02  & 7.6 & 0.18-0.31, 0.11-0.21, 0.05-0.24  \\			
J091205.8+002656 & 2015 Jul 28 & 0.6  & 0.47--0.67 \\		
J091448.7$-$003533 & 2015 Jul 28 & 0.8  & 0.44--0.79 \\	
\hline 
\end{tabular}
\end{center}
\end{table*}

\input{COflux}

\subsection{GBT: {\hi} line}

We performed observations with the Green Bank Telescope (GBT)\footnote{Proposal no.~16A-054 and 16B-037; PI: M.~Micha\l owski.} using the Versatile GBT Astronomical Spectrometer (VEGAS). We used the mode with 61\,kHz spectral resolution (corresponding to 14\,{\kms} at the {\hi} frequency).
The observations were divided into scans lasting 3 or 5 min.
The flux calibration was done using observations of 3C286, whereas pointing and focus was verified using observations of 0744-0629 (radio source \mbox{4C -06.18}) every three hours.
The observing log is presented in Table~\ref{tab:gbt}. 
We used the {\sc GBTIDL} package\footnote{\url{gbtidl.nrao.edu}} to reduce the data. We calibrated each spectrum individually and then averaged them.

We processed the {\hi} GBT/VEGAS spectra in a similar way as for CO.
The data were not usable for J085934.1+003629 and J091205.8+002656 due to strong radio-frequency interference (RFI).
We calculated the atomic gas masses based on eq.~2 in \citet{devereux90}.
The widths of the lines, integrated fluxes, luminosities and the resulting atomic gas masses are presented in Table~\ref{tab:gbtres}.
The {\hi} spectra are shown in the right column in Fig.~\ref{fig:emirspec} in the Appendix. Out of eight targets with usable data  we detected seven.
The features at 500\,{\kms} for J085828.5+003814 and  J090551.5+010752 are likely due to RFI, because they are present only in a fraction of the data.

\begin{table*}
\caption{Observing log for the GBT/VEGAS observations.}
\label{tab:gbt}
\begin{center}
\begin{tabular}{lll}
\hline\hline
Galaxy		& Obs. date 	& $t_{\rm int}$	\\
			&			&  (hr)				  \\
\hline
J085828.5+003814 & 2016 Apr 19, Jul 17, 31, Oct 22 & 10.5    \\		
J085915.7+002329 & 2016 Mar 20 & \phantom{1}3.2    \\		
J085946.7$-$000020 & 2016 Apr 23, 24, Nov 14, Dec 01 & 10.2   \\	
J090038.0+012810 & 2016 Apr 20, Oct 19 & \phantom{1}6.1 	  \\ 		
J090312.4$-$004509 & 2016 Apr 24, 25, Oct 23 & \phantom{1}7.4    \\
J090551.5+010752 & 2016 Mar 21, Jul 16, Oct 18, Nov 19  & \phantom{1}8.3  \\			
J091205.8+002656 & 2016 Nov 20, 27 & \phantom{1}3.5$^a$    \\		
J091448.7$-$003533 & 2016 Mar 12, Oct 23, Nov 29 & \phantom{1}7.2   \\	
J085934.1+003629 & 2015 Apr 18 & \phantom{1}1.8$^a$ \\
J091051.1+020121 & 2016 Mar 19, 20, Nov 01, Dec 01 & \phantom{1}7.4\\
\hline 
\end{tabular}
\tablefoot{$^a$The data not usable due to strong radio-frequency interference.}
\end{center}
\end{table*}

\input{HIflux}

\section{Numerical model}
\label{sec:mod}

To aid the interpretation of the observed data
we use the chemical dust evolution model from \citet[][hereafter GAH11]{gall11}, allowing us to follow the dust and gas removal with stellar age. The model considers SN and asymptotic giant branch (AGB) star dust production at different efficiencies
\citep{gall11b,gall11c}.
It assumes that all material is recycled, and hence available for star formation, instantaneously after the death of the stars. Furthermore, ISM dust destruction through SN shocks is either turned off or set at a moderate rate.    
We calculate a suite of models for which we either switch on or off SN and AGB star dust and gas recycling. Additionally we investigate gas and dust removal through outflows. 

The evolution of dust and gas in a galaxy is: 
\begin{eqnarray}
\label{EQ:DUST}
    \frac{\ud M_{\mathrm{d}}(t)}{\ud t} & = & E_{\mathrm{d,SN}}(t) 
                                        +  E_{\mathrm{d,AGB}}(t)  \nonumber \\
                                        & & {}  -  \eta_{\mathrm{d}}(t) \,
                                        ( \psi(t) + \xi_{\mathrm{SN}}(t) + \zeta_{\mathrm{out}}(t)).               
\end{eqnarray} 
and
\begin{eqnarray}
\label{EQ:GAS}
    \frac{\ud M_{\mathrm{g}}(t)}{\ud t} & = & E_{\mathrm{g,SN}}(t)  +  E_{\mathrm{g,AGB}}(t)    
                                        + \eta_{\mathrm{d}}(t) \, \xi_{\mathrm{SN}}(t)
                                        {}  \nonumber \\
                                        & &  {}  - (1 - \eta_{\mathrm{d}}(t)) \, (\psi(t) +  \zeta_{\mathrm{out}}(t)).
\end{eqnarray}       
Here, $E_{\mathrm{d,SN}}(t)$ and $E_{\mathrm{d,AGB}}(t)$ are the SN and AGB dust production (injection) rates as defined in GAH11 while $E_{\mathrm{g,SN}}(t)$ and $E_{\mathrm{g,AGB}}(t)$ are the rates for the recycled gas phase elements (Eq.~6 in GAH11). The variable $\eta_{\mathrm{d}}(t) = M_{\mathrm{d}}(t) / M_{\mathrm{ISM}}(t)$ can be understood as the  `dust-to-gas mass ratio' or the fraction of dust present in the ISM.  
The evolution of the SFR, $\psi(t)$, here is chosen to represent the measured SFRs (Fig.~1 in \citetalias{michalowski19etg} and eq.~\ref{eq:sfr_age} below) as
\begin{equation}
\label{EQ:SFR}
         \psi(t) = \psi_{\mathrm{ini}} \, 
         e^{{-} \, a(t)}, \, a(t) = \frac{ T_{\mathrm{gal}}(t)}{\tau_{\mathrm{SFR}}},
         \label{eq:sfr_age_model}
\end{equation}
where $\psi_{\mathrm{ini}}$ is the initial SFR, $T_{\mathrm{gal}}(t)$ is the age of the galaxy and $\tau_{\mathrm{SFR}}=1.8\,$Gyr 
is the star formation decline timescale (duration of the star formation episode). We note that this SFR evolution is not coupled to the total amount of gas in the galaxy as it would be by a Schmidt-Kennicutt law \citep{kennicutt98} and as defined in GAH11. Instead, the formulation here is based on observations (Fig.~1 in \citetalias{michalowski19etg}) of the galaxies in question.  

An additional cold gas removal is considered, either in a form of gas heating or physical ISM removal. It is modeled as
\begin{equation}
\label{EQ:OUTF}
          \zeta_{\mathrm{out}}(t) = \frac{M_{\mathrm{ISM, out}}}{\tau_{\mathrm{ISM}}} \, 
         e^{{-} \, b(t)}, \, b(t) = \frac{ T_{\mathrm{gal}}(t)}{\tau_{\mathrm{ISM}}},
\end{equation}
where $M_{\mathrm{ISM, out}}$ is the amount of ISM material to be removed over a timescale $\tau_{\mathrm{ISM}}$.

The variable $\xi_{\mathrm{SN}}(t) = {M_{\mathrm{cl}} \, R_{\mathrm{SN}}(t)}$ defines how much ISM mass, $M_{\mathrm{cl}}$, is completely cleared of dust by core-collapse SNe 
\citep[CCSNe;][]{dwek07},
with $R_{\mathrm{SN}}(t)$ being the CCSN rate.
We have been testing zero to moderate dust destruction with $M_{\mathrm{cl}}$ of 0, 50, 100 or 500 \msun.
Both the CCSNe rate and AGB rate are calculated as
\begin{equation}
\label{EQ:SNAGBRAT}
         R_{\mathrm{AGB, SN}}(t) = \int_{m_{\mathrm{L(AGB, SN)}}}^{m_{\mathrm{U(AGB, SN)}}}
                                                \psi(t - \tau) \,
                                                 \phi(m) \, \ud m   ,  
\end{equation}
which assumes that only single stars form. $R_{\mathrm{AGB, SN}}(t)$ is regulated by the IMF, $\phi(m)$, and $\psi(t - \tau)$ is the lifetime of a star with a given zero-age main sequence mass $m$. 
The model parameters are summarized in Table~\ref{tab:demonpar}.

\section{Results}
\label{sec:res}

\begin{figure*}
\includegraphics[width=0.8\textwidth]{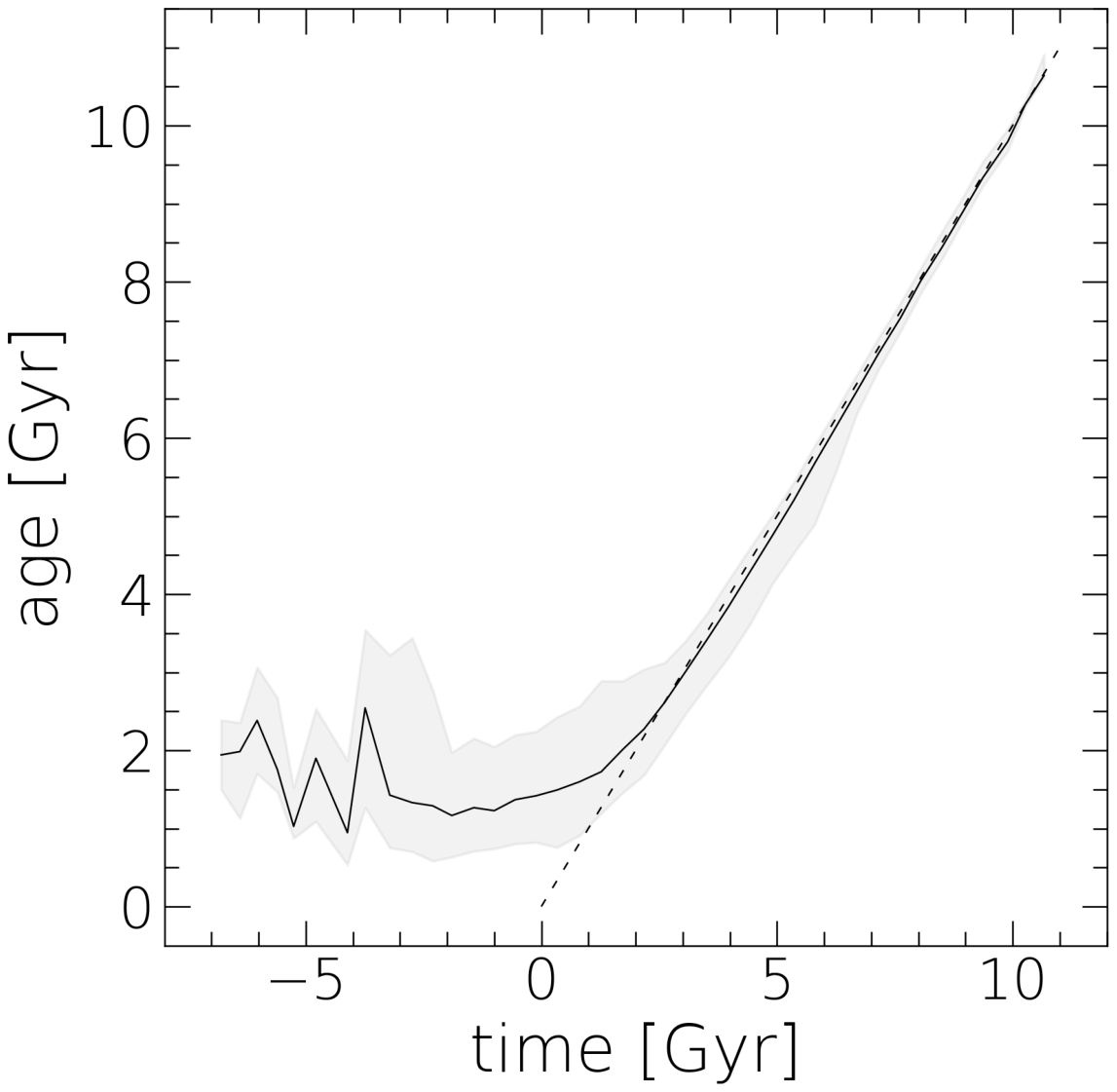}
 \caption{Light-weighted stellar age as a function of cosmic time for simulated dusty ETGs (Nadolny et al., in prep.). The {\it black solid line} is the running average for given time, whereas the {\it grey shaded region} represent the $1\sigma$ range. The dashed line indicates the linear evolution when the stellar age is equal to the value of cosmic time. The time reaches negative values because of how the moment when time is equal zero was set, in order to have the time equal to the stellar age at the moment of observation. This figure shows that stellar ages increase linearly with time for dusty ETGs, so can be used as a time proxy.
}
 \label{fig:age_time}
\end{figure*}

Our aim is to investigate the time evolution of various galaxy properties, using the stellar age as a time proxy, so first we verify that indeed the ages of these galaxies increase linearly with time. In principle this is not the case for all galaxies, especially for highly star-forming ones for which the mean stellar age may even decrease with time. Strictly speaking the stellar age increases linearly with time when star formation rate is zero. Our galaxies have very low levels of star formation and large stellar masses, so they should be close to this approximation, because the mean stellar age is not affected significantly by the presence of  new populations of stars and therefore should grow linearly with time.

We checked the age evolution with time using 2\,000 simulated analogs of our galaxies selected in a very similar way: similar redshift distribution and stellar masses, dust masses, so that the $250\,\micron$ flux would be detectable as for the real sample, early-type morphological classification, and low star-formation activity placing them below the main-sequence (Nadolny et al., in prep.). 
These are Millennium Simulations with LGalaxies semi-analytical models with $\sim100^3\,\mbox{Mpc}^3$ box \citep{lemson06,springel05,henriques20}. Unlike for observed galaxies, for each of the simulated galaxies we have a complete knowledge of the light-weighted stellar age as a function of the Universe age (cosmic time). 

In order to analyse the time evolution of ages for the entire sample, we need to choose a common time zeropoint. Galaxies observed at similar ages of the Universe have a range of stellar ages, so in order to assess how the stellar age evolve with time for the entire sample, we shifted horizontally (in time) the age-time tracks, so that at the time of observation the cosmic time is equal to the measured stellar age. With such a choice of the time zeropoint, galaxies with higher levels of current star formation activity than in the past should be located below the $\mbox{age}=\mbox{time}$ line (ages lower than the cosmic time value) due to a significant number of young stars. On the other hand, galaxies with long periods of constant star formation rate should be located above this line (ages higher than the cosmic time value), because their mean stellar age does not change in time. 

Fig.~\ref{fig:age_time} shows the age-time evolution for simulated dusty ETGs. It is clear that neither of the non-linear scenarios described above applies to them. Their mean stellar ages increase linearly with time at least during the last 8\,Gyr, which is the range of interest of our study. Very low scatter around the $\mbox{age}=\mbox{time}$ line indicates that very few galaxies exhibit levels of star formation strong enough to break the linearity of stellar ages with time.

\begin{figure*}
\includegraphics[width=0.8\textwidth]{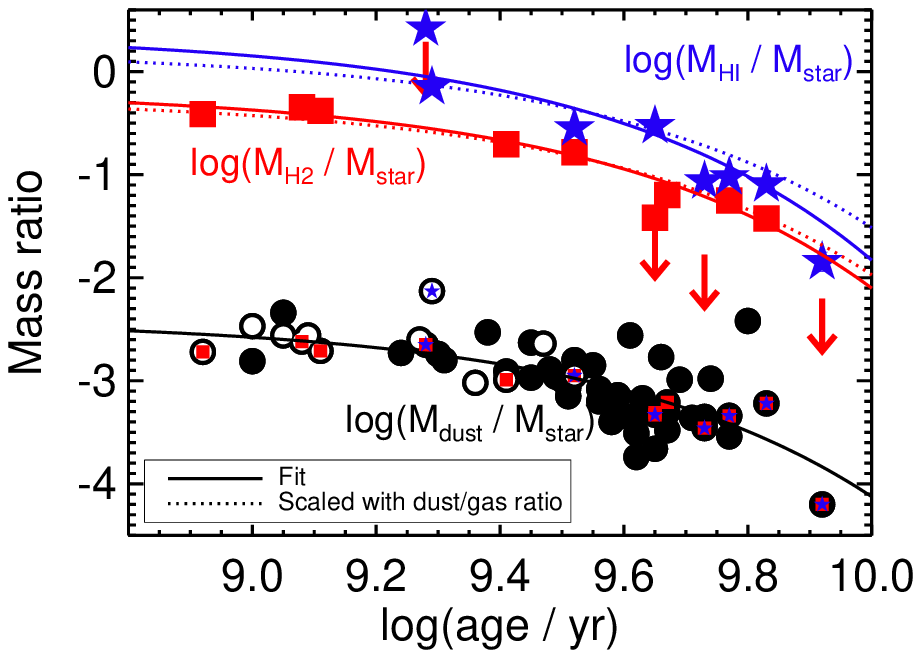}
 \caption{Ratios of molecular ({\it large red squares} and {\it arrows} for upper limits), atomic ({\it large blue stars}) gas, and dust ({\it black circles}; \citetalias{michalowski19etg}) masses to stellar masses as a function of luminosity-weighted stellar age of ETGs galaxies detected by {\it Herschel} from the sample of \citet{rowlands12}. The exponential fits to the gas-to-stellar and dust-to-stellar mass ratios 
(eq.~\ref{eq:mgasms_age} and \ref{eq:mdms_age})
are shown as {\it solid lines}, colored as the datapoints.
 The {\it red} and {\it blue dotted lines} denote the curve for the dust-to-stellar ratio shifted upwards by the measured average gas-to-dust ratios of $\log(\mhtwo/\mdust)= 2.144\pm0.018$ and $\log(\mhi/\mdust)=2.610\pm0.011$.
{\it Small red squares} and {\it blue stars} mark the galaxies which we observed at CO and {\hi}, respectively.
Open circles denote galaxies which are within the main sequence (see Fig.~1 in \citetalias{michalowski19etg}).
All ISM components decline at a similar rate.
}
 \label{fig:mdms_age}
\end{figure*}

Having demonstrated that the stellar ages of dusty ETGs increase linearly with time, at least for the dusty ETGs selected in the way we do here, we come back to the observed sample.
In order to compare galaxies with different masses, we normalize the gas and dust masses 
by their stellar masses.
We show the molecular and atomic  gas-to-stellar mass ratios ($\mhtwo/\mstar$, $\mhi/\mstar$) as a function of stellar age in Fig.~\ref{fig:mdms_age}. We detect a decline of the gas-to-stellar mass ratios with age, similar to the  evolution of the dust-to-stellar ratio ($\mdust/\mstar$) from \citetalias{michalowski19etg} and \citet{lesniewska23}.
For the $\mhtwo/\mstar$--age and $\mhi/\mstar$--age diagrams the Spearman rank correlation coefficients are $-0.9$ and the   probabilities of the null hypothesis of no correlation are $0.00094$ ($3.3\sigma$) and $0.019$ ($2.3\sigma$), respectively.
Due to a smaller sample size, this significance is lower than $5.5\sigma$ we reported for the correlation between $\mdust/\mstar$ and age (probability of $4\times10^{-11}$; \citetalias{michalowski19etg}). 

Fig.~\ref{fig:mdms_age} presents the decline of ISM mass as a function of age.
We fitted an exponential function to the gas-to-stellar mass ratios and obtained
\begin{eqnarray}
    \log(\mhtwo/\mstar)=(-\mbox{age}/\mbox{Gyr})/(5.21\pm0.24) - (0.18\pm0.03) \nonumber  \\
    \log(\mhi/\mstar)=(-\mbox{age}/\mbox{Gyr})/(4.54\pm0.23) + (0.37\pm0.05)\nonumber  \\
    \label{eq:mgasms_age}
\end{eqnarray} 

This corresponds to the characteristic exponential timescale of gas removal of $\tau_{H_2} = 2.26 \pm   0.11$\,Gyr and $\tau_{\rm HI} = 1.97 \pm   0.10$\,Gyr
(half-life time of  $t_{1/2\,H_2}=\tau_{H_2}\ln{2}=1.57 \pm   0.07$\,Gyr and $t_{1/2\,{\rm HI}}=1.37 \pm   0.07$\,Gyr).
The gas removal timescales are consistent with the dust removal timescales (at $2\sigma$) we measured in \citetalias{michalowski19etg} and \citet{lesniewska23}, where we obtained the  characteristic exponential timescale of dust removal to be $\tau_{\rm dust} = 2.53 \pm   0.17$ and $2.26\pm0.18$\,Gyr, respectively
(half-life time of  $1.75 \pm   0.12$ and $1.57\pm0.12$\,Gyr).

The decline of the gas amount in ETGs with time was seen in models \citep{calura17}, but has not been measured directly before. On the other hand, \citet{smercina18} detected a dust decline with age for post-starburst galaxies with ages up to 1\,Gyr, and a very weak gas decline. They interpret this as the effect of sputtering of dust grains in hot gas.
With a larger sample of post-starburst galaxies, \citet{french18} detected a gas decline on the timescale of 100--200\,Myr.
Similarly, molecular gas was found in post-starburst galaxies at $z\sim0.6$ only for those which were quenched less than 150\,Myr ago, indicating a rapid gas removal \citep{bezanson22}.
The timescale we measure is comparable to that inferred by \citet{gobat20}, based on the low gas fraction of low-$z$ ETGs, and higher than the quenching timescale due to environmental influence \citep{gobat15}.

The weighted average of the gas-to-dust ratios are $\log(\mhtwo/\mdust)= 2.144\pm0.018$ and $\log(\mhi/\mdust)=2.610\pm0.011$.
In Fig.~\ref{fig:mdms_age} we used these averages to scale up the exponential fit to the dust-to-stellar ratios (\citetalias{michalowski19etg}):
\begin{equation}
    \log(\mdust/\mstar)=(-\mbox{age}/\mbox{Gyr})/(5.8\pm0.4) - (2.41\pm0.09)
    \label{eq:mdms_age}
\end{equation} 
Indeed the evolution of the gas-to-stellar ratios are consistent with these scaled curves.

\begin{figure*}
\includegraphics[width=0.8\textwidth]{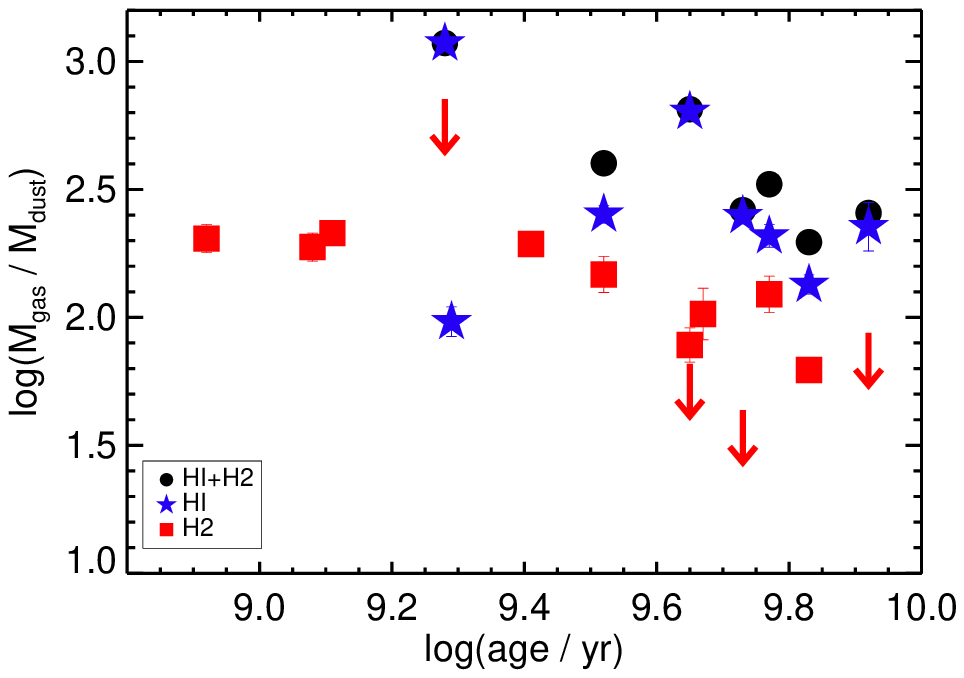}
 \caption{Gas-to-dust ratios as a function of luminosity-weighted stellar age. {\it Red squares}, {\it blue stars}, and {\it black circles} denote molecular gas, atomic gas, and total gas, respectively (the latter only for galaxies with both CO and {\hi} measurements). 
 Only very week trends are present indicating that the ISM components are affected by the same mechanism.
}
 \label{fig:mh2md_age}
\end{figure*}

The gas-to-dust ratios are either independent of or weakly declining with age (Fig.~\ref{fig:mh2md_age}). There is a hint of a decline of the $\mhtwo/\mdust$ ratio with age at a $\sim2.9\sigma$ level (probability of the null hypothesis of no correlation of $0.0037$  and the  Spearman rank correlation coefficient  of $-0.85$). There is no indication of the decline of the $\mhi/\mdust$ ratio (the  Spearman rank correlation coefficient  of $-0.09$, the probability of the null hypothesis of $0.87$).
We note that the gas-to-dust ratios we found are similar to those of star-forming galaxies. Hence, these ETGs do not belong to the class of passive galaxies at $0<z<3$ with extremely high ratios, recently identified in simulations \citep{whitaker21,donevski23}.

\begin{figure*}
\includegraphics[width=0.8\textwidth]{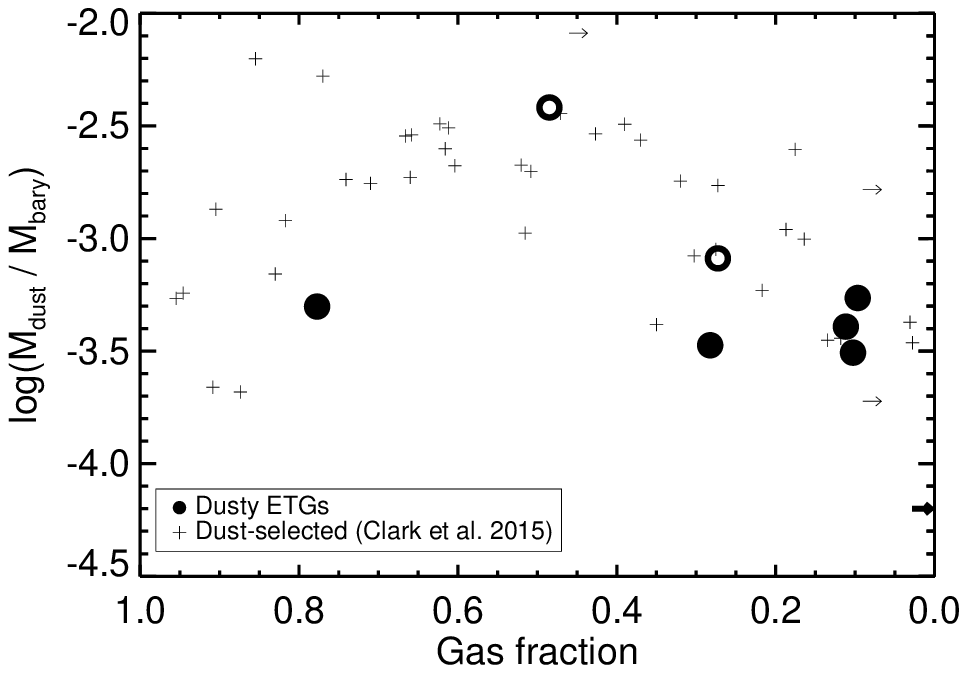}
 \caption{Dust-to-baryon mass ratio [$\mdust/(\mstar+\mhi)$] as a function of atomic gas fraction [$\mhi/(\mstar+\mhi)$] of ETGs in our sample ({\it large black circles and a thick arrow}) compared with dust-selected galaxies from \citet[][{\it plus signs and thin arrows}]{clark15}. {\it Open circles} denote galaxies which are within the main sequence see Fig.~1 in \citetalias{michalowski19etg}).
 The ETGs in our sample are located in the low gas fraction regime, expected for old galaxies.
}
 \label{fig:mdmb_fg}
\end{figure*}

The ETGs in our sample follow the evolutionary sequence found for other galaxies based on ISM properties. In Fig.~\ref{fig:mdmb_fg} we show the dust-to-baryon mass ratio [$\mdust/(\mstar+\mhi)$] as a function of atomic gas fraction [$\mhi/(\mstar+\mhi)$] together with dust-selected galaxies \citep{clark15}. Modelling shows that galaxies evolve from left (high gas fractions) to right (low gas fractions), increasing their dust-to-baryon ratios initially and then following a decline \citep{clark15,devis17,devis17b,devis19,donevski20,nanni20}. The ETGs in our sample have low gas fractions, consistently with being among the oldest galaxies in this plot.
The only galaxy with high gas fraction is J085915.7+002329, having the lowest stellar mass in our sample [$\log(\mstar/\msun)=8.62$]. It is indeed one of the youngest as well with an age of $10^{9.3}$\,yr.

\begin{figure*}
\includegraphics[width=0.8\textwidth]{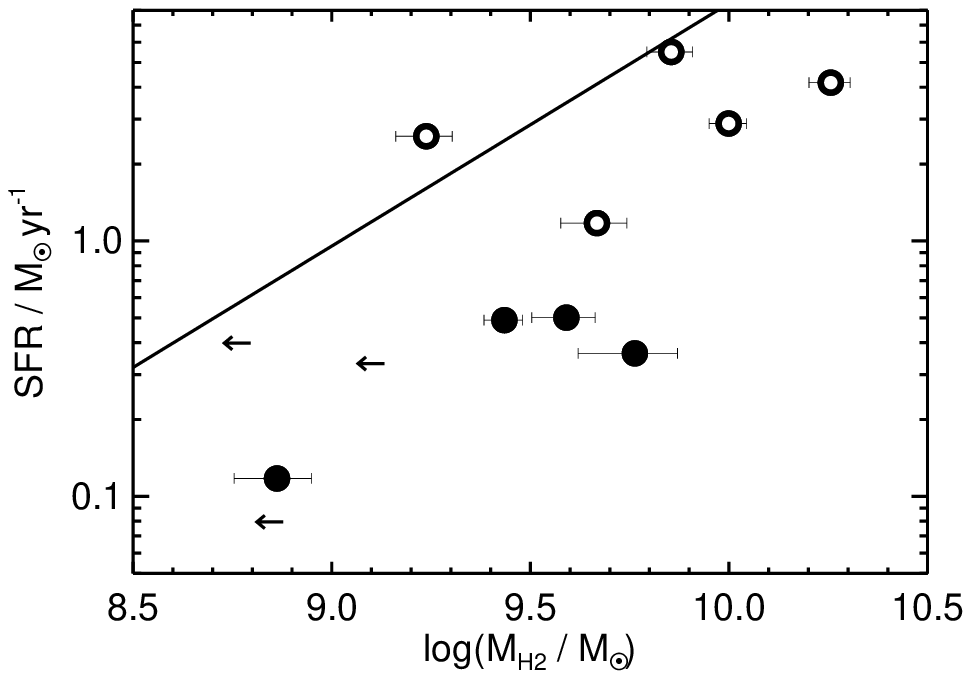}
 \caption{Star formation rates as a function of molecular gas masses.
 Arrows denote upper limits. Open circles denote galaxies which are within the main sequence (see Fig.~1 in \citetalias{michalowski19etg}). The {\it solid line} denotes the relation for star-forming galaxies \citep[eq.~1 in][]{michalowski18co}. 
 ETGs are below the relation for star-forming galaxies, indicating that their gas reservoir is ceasing to be able to form stars.
}
 \label{fig:mh2_sfr}
\end{figure*}

\begin{figure*}
\includegraphics[width=0.8\textwidth]{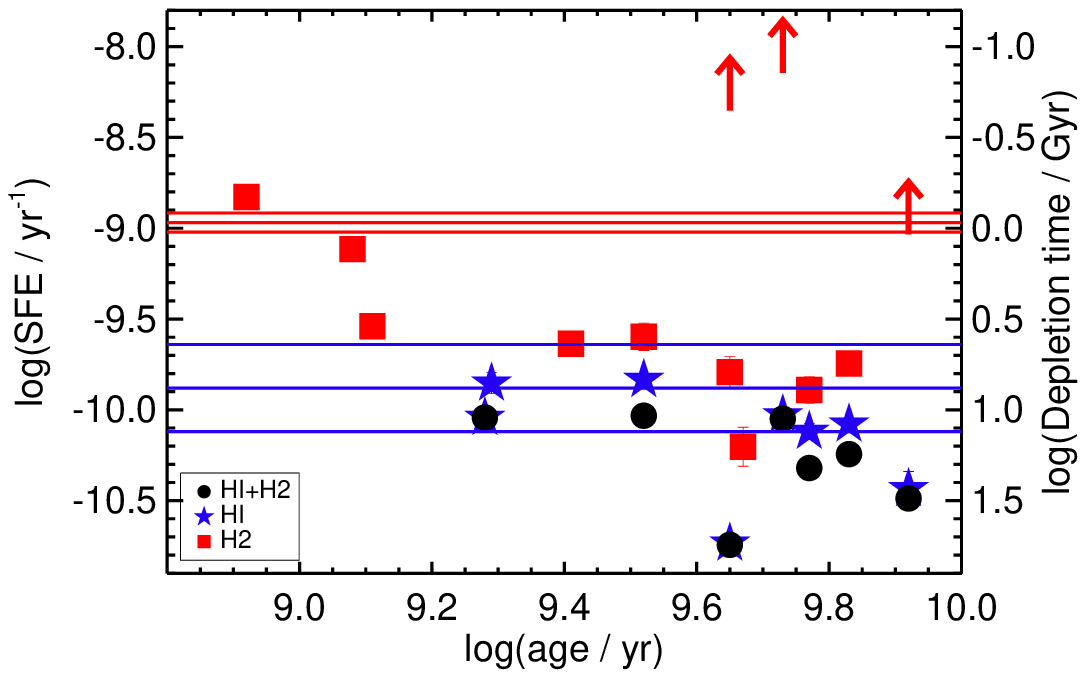}
 \caption{Star formation efficiency ($\mbox{SFE}\equiv\mbox{SFR}/\mgas$) as a function of luminosity-weighted stellar age.
 {\it Red squared}, {\it blue stars}, and {\it black circles} denote molecular gas, atomic gas, and total gas, respectively (the latter only for galaxies with both CO and {\hi} measurements).
 The solid lines represent the SFE of star forming galaxies with $\mbox{SFR}=0.01$, $0.1$, and $1\msunyr$ (from bottom to top) using eq.~1 in \citet[][atomic gas]{michalowski15hi} and eq.~1 in \citet[][molecular gas]{michalowski18co} colored in the same way as datapoints.
 The molecular SFEs decline with age and are lower than those of star-forming galaxies, whereas the atomic SFEs stay consistent with those of star-forming galaxies.
}
 \label{fig:sfe_age}
\end{figure*}

The SFRs of the ETGs in our sample decline faster than the mass of the ISM, as noted in \citet{hjorth14}. The exponential fit of \citetalias{michalowski19etg} (their Fig.~1) results in:
\begin{equation}
    \log(\mbox{SFR})=(-\mbox{age}/\mbox{Gyr})/(4.0\pm0.4) + (0.57\pm0.09)
    \label{eq:sfr_age}
\end{equation} 
which corresponds to the characteristic exponential time $\tau_{\rm SFR} =  1.8 \pm   0.4$\,Gyr or a half-life time $t_{1/2,{\rm SFR}}= 1.22 \pm   0.28$\,Gyr. This is $\sim40\%$ faster than the dust decline (eq~.\ref{eq:mdms_age}).

We note that the exponential decline parametrization for SFR in the above equation and in the model (eq.~\ref{eq:sfr_age_model}) is justified for these old galaxies. First, this function fits the data (Fig.~1 of \citetalias{michalowski19etg}). Second, the SED fits from \citet{rowlands12} resulted in no recent starbursts in this sample with the time since the last starburst $>500$\,Myr for all galaxies, $>1$\,Gyr for 80\% of the sample, and $>2$\,Gyr for 56\% of the sample. The existence of these bursts does not affect our analysis because in eq.~\ref{eq:sfr_age} we quantify the evolution of the average SFR after these episodes. Similarly, the assumption of the exponential decline of the SFR in the model has almost no impact on the resulting properties, because the SFRs are low, so the astration is weak, independent of what parametrization is adopted, as long as it is consistent with the low measured values.

Galaxies in our sample have low SFRs for their molecular gas masses, as demonstrated in Fig.~\ref{fig:mh2_sfr}. They are below the Schmidt-Kennicutt law by a factor of several.


Hence, the star formation efficiency ($\mbox{SFE}\equiv\mbox{SFR}/\mgas$) of our sample is low with a mean value of $\log(\mbox{SFE}/\mbox{yr}^{-1})\sim-9.6$, $-10.2$ and $-10.3$ (depletion times $1/\mbox{SFE}\sim4$, $16$, and $20$\,Gyr), using molecular, atomic, and total gas mass, respectively (Fig.~\ref{fig:sfe_age}). These efficiencies are similarly low as for the sample of ETGs with dust lanes \citep{davis15} and some post-starburst galaxies \citep{alatalo15b,french15,french23,rowlands15,smercina18,smercina22,bezanson22,luo22}; at the lower limit for high-$z$ ETGs \citep{magdis21}; and lower than the full ETG ATLAS$^{\rm 3D}$ sample \citep{davis14, kokusho17}. In Fig.~\ref{fig:sfe_age} we compare the SFEs of our sample with those of the general star formation population. For molecular and atomic gas we used equation~1 of \citet{michalowski18co} and  equation~1 of \citet{michalowski15hi}, respectively \citep[see similar estimates in][]{sargent14,saintonge17,liu19,magdis21}. Again, the molecular SFEs of galaxies in our sample are lower than those of star-forming galaxies.
The difference is stronger for older ages, so these galaxies keep decreasing their efficiency with time. However, the atomic SFEs are similar to those of star-forming galaxies. This could be because the atomic gas in ETGs is only indirectly connected with star formation.

\begin{figure*}
\includegraphics[width=0.8\textwidth]{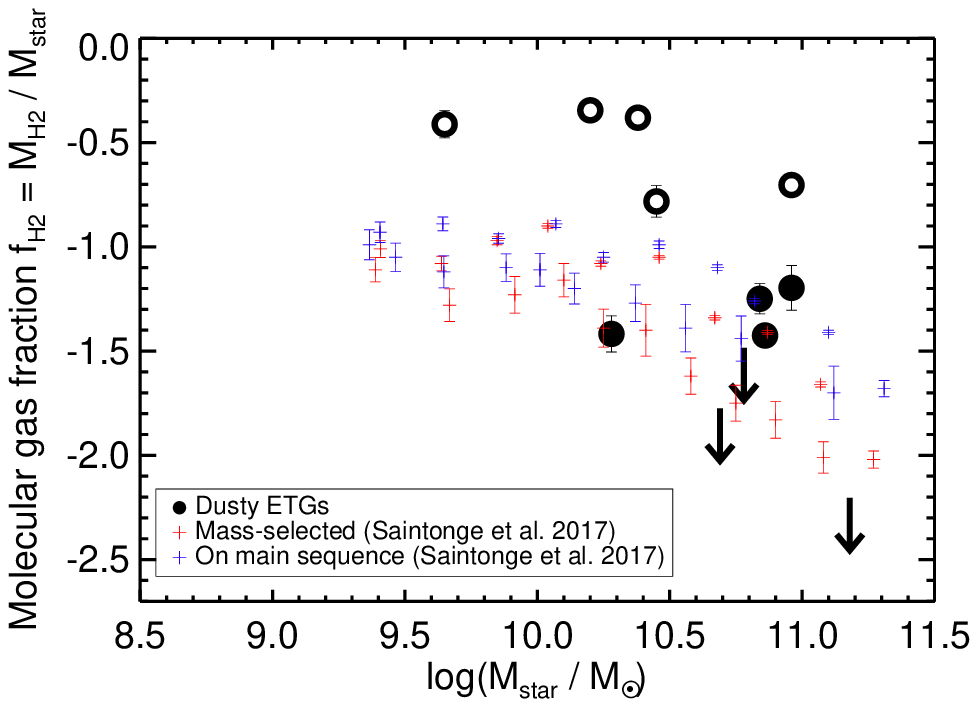}
 \caption{Molecular gas fraction ($\fhtwo\equiv \mhtwo/\mstar$) as a function of stellar mass for the sample of ETGs in our sample (black circles and arrows). Open circles denote galaxies which are within the main sequence (see Fig.~1 in \citetalias{michalowski19etg}). The red and blue symbols correspond to averages of mass-selected galaxies and only those on the main sequence, respectively \citep{saintonge17}.
 Gas fractions of ETGs are comparable or higher than those of star-forming galaxies, indicating that the ETGs do not stop forming stars due to lack of gas.
}
 \label{fig:fh2_ms}
\end{figure*}

\section{ISM removal and quenching}
\label{sec:removal}

We have provided the 
a measurement of the timescale of ISM removal in ETGs of $\sim2.3$\,Gyr 
(eq.~\ref{eq:mgasms_age}).
This slow ISM removal suggests that either quenching in these galaxies is a slow process or that the main reason for quenching is not exhaustion of the gas supply, but rather gas stabilization that prevents further star formation.
Indeed, we have measured a slow SFR decline (eq.~\ref{eq:sfr_age}).
Similarly, the timescale of quenching was measured to be of the order of several Gyr in observed \citep{peng15,trussler20,kipper21,tacchella22,noirot22,bravo23,donevski23} and simulated \citep[][Nadolny et al., in prep.]{trayford16,nelson18,wright19,park22,walters22} galaxies, with the exception of cluster members in which it is shorter \citep{muzzin14,socolovsky18,davis19c,zavala19}.

As detailed below, the galaxies in our sample are shutting down their star formation not only because they are running out of gas, but because their SFRs decline faster than the gas amount and they have low SFEs and normal gas fractions.

Even in samples of post-merger and post-starburst galaxies the gas supply was actually higher than in other galaxies with similar masses \citep{rowlands15,french15,alatalo16,suess17,ellison18}, so quenching in these galaxies is likely related to turbulence, not exhaustion or expulsion of gas. Morphological quenching, the influence of the bulge making the gas resilient against fragmentation \citep{martig09,martig13,bluck14,bluck20b,bitsakis19,lin19,gensior20} or the influence of turbulence and magnetic fields \citep{padoan02,federrath12} can also be responsible for decreasing SFRs.
Indeed galaxies in our sample evolve upwards from the SFR--{\mdust} relation (having higher {\mdust} than what their SFRs would imply), as predicted for quenching that is not caused by the removal of gas 
\citep[][\citetalias{michalowski19etg}]{hjorth14,lesniewska23}.
However, in simulations morphological quenching is found to be effective at gas fractions below a few \% \citep{martig13,gensior20}, much lower than for our sample (Fig.~\ref{fig:fh2_ms}).

The very low SFE of our ETGs (Fig.~\ref{fig:sfe_age}) indicate that star formation is suppressed even in comparison with other ETGs.
This means that the process shutting down the SFRs in these galaxies is not due to physical gas removal, but to its inability to form stars. 
This again supports the internal quenching scenario, either morphological or connected with turbulence or magnetic fields.

For the ETGs in the ATLAS$^{\rm 3D}$ sample, \citet{kokusho17}  found no decline in SFE as a function of age, and found  consistency with the SFR-{\mhtwo} relation of star-forming galaxies. Hence, together with the decline in the gas fraction with age, this was interpreted as quenching driven by a decrease of the gas reservoir.  In contrast, we do see a decline in the SFE in our sample (Fig.~\ref{fig:sfe_age}) and they are below the SFR-{\mhtwo} relation (low SFRs for their molecular gas masses; Fig.~\ref{fig:mh2_sfr}), again pointing at quenching being connected with the decreased ability of gas to form stars, not with a lack of gas.

Moreover, the high molecular gas fractions also point to declining gas amount not being the main reason for declining SFRs. Fig.~\ref{fig:fh2_ms} shows the molecular gas fraction ($\fhtwo\equiv \mhtwo/\mstar$) as a function of stellar mass compared with mass-selected low-redshift galaxies \citep{saintonge17}. The ETGs in our sample on the main sequence (lower ages) exhibit larger gas mass fractions than these star-forming galaxies, whereas the ETGs below the main sequence have comparable gas fractions to star-forming galaxies.  Hence the lack of gas is  not the main reason for the ETGs to become passive. This is different from green-valley galaxies, for which both SFE and {\fhtwo} were found to be suppressed \citep{brownson20,lin22}.

We note that our galaxies may not be representative of the entire ETG population, because they were selected based on {\it Herschel} $250\,\micron$ detections. This corresponds to the dust mass detection threshold of $10^{5.2}\,\msun$ at $z=0.05$ and $10^{6.7}\,\msun$ at $z=0.3$ \citepalias{michalowski19etg}. These are not very high limits, especially given the high stellar masses of our galaxies, but most ETGs were shown to have less dust. In particular, in the ETG parent sample of \citet{rowlands12}, only 5.5\% were detected by {\it Herschel}. The remaining dust-poor ETGs may follow different evolutionary tracks than those presented here.

\section{Mechanism of ISM removal}
\label{sec:mechanism}



Here we discuss the possible physical mechanisms explaining the trend depicted in Fig.~\ref{fig:mdms_age}. For each mechanism we list the predictions which can then be compared with existing and future datasets.
The predictions are summarized in Table~\ref{tab:pred}.
Only two mechanisms are fully consistent with all the current data: removal of the entire cold ISM (Section~\ref{sec:ismdes}) and outflows (Section~\ref{sec:outflow}). They are not mutually exclusive, so it could be that they operate together. In Section~\ref{sec:esource} we also discuss the energy source required for these mechanisms.

\begin{table*}
\caption{Predictions of the mechanisms that can explain dust decline in Fig.~\ref{fig:mdms_age} and their consistency with the currently available data.  \label{tab:pred}}
\begin{tabular}{lll}
\hline\hline
Mechanism & Prediction & Agreement \\
		&			& with data\\
\hline
Removal of the entire cold ISM (\ref{sec:ismdes}) & constant or decreasing $M_{\rm gas}/\mdust$ & $\checkmark$, Fig.~\ref{fig:mh2md_age} \\
	& {\mstar} constant or slightly increasing with age & $\checkmark$, 
	Fig.~1 in \citetalias{michalowski19etg}\\
	& uniform or centrally concentrated ISM distribution & ? \\
\hline
Outflows (\ref{sec:outflow}) & constant or decreasing $M_{\rm gas}/\mdust$ & $\checkmark$, Fig.~\ref{fig:mh2md_age}\\
	& {\mstar} constant or slightly increasing with age & $\checkmark$, 
	Fig.~1 in \citetalias{michalowski19etg}\\
	& off-center filamentary or plume-like gas structures & ?\\
	& multiple velocity peaks or broad line wings & ?, Fig.~\ref{fig:emirspec}\\
\hline
Dust grain destruction (\ref{sec:graindes}) & $M_{\rm gas}/\mdust$ increasing with age & X, Fig.~\ref{fig:mh2md_age} \\
	& {\mstar} constant or slightly increasing with age & $\checkmark$, 
	Fig.~1 in \citetalias{michalowski19etg}\\
\hline
Astration/strangulation (\ref{sec:starv}) &  constant $M_{\rm gas}/\mdust$ & X, Fig.~\ref{fig:mh2md_age} \\
	& {\mstar} constant or slightly increasing with age & $\checkmark$, 
	Fig.~1 in \citetalias{michalowski19etg}\\ 
	& $\mdust/{\mstar}$ and $M_{\rm gas}/{\mstar}$ flattening at higher ages & X, Fig.~\ref{fig:mdms_age} \\
\hline
Decreasing number of  AGB stars (\ref{sec:agbnumber}) & AGB stars dominate dust production & X, \ref{sec:agbnumber} \\
\hline
Dust cooling (\ref{sec:cool}) & {\submm} excess & X, Fig.~5 in \citetalias{michalowski19etg} \\ 
	& dust temperature decreasing with age & X, Fig.~1 in \citetalias{michalowski19etg} \\ 
	& $M_{\rm gas}/\mdust$ increasing with age & X, Fig.~\ref{fig:mh2md_age} \\
\hline
Dust heating (\ref{sec:heat}) & SEDs peaking at short wavelengths & X, Fig.~A1 in \citetalias{rowlands12}\\
	& dust temperature increasing with age & X, Fig.~1 in \citetalias{michalowski19etg}\\ 
	& $M_{\rm gas}/\mdust$ increasing with age & X, Fig.~\ref{fig:mh2md_age} \\
\hline
Environmental influence (\ref{sec:env}) & environmental density increasing with age & X, Fig.~1 in \citetalias{michalowski19etg} \\ 
	& rich environments & X, Fig.~1 in \citetalias{michalowski19etg} \\ 
\hline
Mergers with gas-rich galaxies (\ref{sec:merger}) & $M_{\rm gas}/\mdust$ increasing with the derived age & X, Fig.~\ref{fig:mh2md_age} \\
	& {\mstar} decreasing with the derived age & X, 
	Fig.~1 in \citetalias{michalowski19etg}\\
	& $M_{\rm gas}/{\mstar}$ decreasing with the derived age & $\checkmark$, Fig.~\ref{fig:mdms_age} \\
	& $\mdust$ only weakly correlated with {\mstar} & X, Fig.~3 in \citetalias{michalowski19etg} \\ 
	& size increasing with decreasing age & X, Fig.~2 in \citetalias{michalowski19etg}\\ 
 \hline
$\mstar$-age correlation (\ref{sec:magecorr}) & no $\mdust/\mstar$-age anti-correlation  & X, Fig.~\ref{fig:mdms_age_narrowms}\\
& for narrow ranges of $\mstar$\\
\hline
Selection bias (\ref{sec:bias}) & bias against old ISM-rich galaxies  & X\\
                                        &  bias against young ISM-poor galaxies  & X\\
\hline
\end{tabular}
\tablefoot{$\checkmark$: the prediction is consistent with the data, X: the prediction is inconsistent with the data, ?: the data needed to test this prediction is not available yet. \citetalias{rowlands12}: \citet{rowlands12}, \citetalias{michalowski19etg}: \citet{michalowski19etg}.}
\end{table*}

\subsection{Removal of the entire cold ISM}
\label{sec:ismdes}

The destruction of all components of the cold ISM 
involves destroying dust, molecular and atomic gas.
We will consider mechanisms having a stronger effect on dust in Section~\ref{sec:graindes}.
There may be several physical mechanisms responsible for the removal of the entire ISM (planetary nebulae, SNe Type Ia, AGN, hot gas, cosmic rays). 
 We will return to discussing these energy sources in Section~\ref{sec:esource}.

The gas which is ceasing to be in the cold (molecular or atomic) phase is transformed into ionized hot gas. In our sample this requires heating a few times $10^9\,\msun$ of gas. 
The hot gas in elliptical galaxies has a mass corresponding to 1\% of stellar mass (or 10-20\% for the most massive ones;
\citealt{mathews03,sparke06}). 
For our galaxies with stellar masses of $\mstar=10^{10-11}\,\msun$, this corresponds to $10^{8-9}\,\msun$ of hot gas (or more for the most massive galaxies in our sample).
Hence, the amount of gas needed to be ionized by this mechanism is similar to the typical hot gas reservoirs in such galaxies, taking into account that some of the gas will be expelled or used for star formation.

This scenario predicts the following. 

\begin{enumerate}
\item $M_{\rm gas}/\mdust$ should be constant, as both dust and gas are destroyed, or decreasing with age if gas particles are destroyed faster due to their more diffuse distribution than dust, which has a more clumpy distribution.
\item {\mstar} should be constant or slightly increasing with age, given the low SFRs of galaxies in our sample.
\item Gas and dust should be relatively uniformly distributed with possible either central concentration, reflecting the initial distribution before the quenching, if the process removing the cold ISM is not violent and operates throughout galaxies, or central deficit, if the source of the energy is in the galaxy center (AGN).
\end{enumerate}

The $M_{\rm gas}/\mdust$ ratio indeed declines slightly with age  (Fig.~\ref{fig:mh2md_age}) and {\mstar} is slightly increasing 
(Fig.~1 in \citetalias{michalowski19etg}).

Our numerical model (Section~\ref{sec:mod}, Fig.~\ref{fig:mdms_age_model}, Table~\ref{tab:demonpar}) shows that in order to explain the data, around $3$--$5\times10^{10}\,\msun$ of gas (including both the atomic and molecular phase) needs to be removed on the timescale of 10\,Gyr (the average rate of $3$--$5\,\msunyr$, which includes the phase of being a normal star-forming galaxy). We note that this is much higher than the measured SFRs for higher ages, at which the gas decline is the strongest, so astration is unlikely to dominate the ISM mechanism (see Section~\ref{sec:starv}).

\subsection{Outflows}
\label{sec:outflow}

Gas and dust can be expelled from galaxies either by AGN- or stellar-induced winds. 
Outflows have been detected in M82 \citep{walter02}, Arp\,220 \citep{perna20}, other starbursts \citep{tsai09,tsai12,bolatto13b,cicone14,geach14b,geach14,walter17b,maiolino17}, and ETGs \citep{alatalo11} and are present in simulations \citep{dallavecchia12,vogelsberger14,schaye15,dave19,nanni20,burgarella20}. Outflows can also be driven by cosmic-rays \citep{hopkins21b}. This mechanism has been proposed as being responsible for shutting down star-formation, at least in the most massive galaxies.

This scenario predicts the following.

\begin{enumerate}
\item $M_{\rm gas}/\mdust$ should be constant with age, as both components are being removed, or decreasing with age if dust distribution is more clumpy (as in Section~\ref{sec:ismdes}). \label{pred:out:gd}
\item {\mstar} should be constant or slightly increasing with age, given low SFRs of galaxies in our sample. \label{pred:out:mstar}
\item Some molecular gas should be located away from optical 
 centers
of the galaxies, possibly in
filamentary or plume-like structures with scales of around 1--2\,kpc \citep{walter02,feruglio10,feruglio15,alatalo11,morganti15}. \label{pred:out:distr}
\item Multiple velocity peaks with locations incompatible with a rotating disk, or broad line wings should be present \citep[e.g.][]{cicone14}. \label{pred:out:spec}
\end{enumerate}

Predictions \ref{pred:out:gd} and \ref{pred:out:mstar} are consistent with the data. We do not have spatial information on the distribution of gas to test prediction \ref{pred:out:distr}. In principle we can look for high-velocity wings in the spectra (Fig.~\ref{fig:emirspec}, prediction \ref{pred:out:spec}), but the expected signal is at the level of 10\% of the main CO line peak \citep[e.g.][]{cicone14}. Our data are not of enough signal-to-noise ratio to test this.

We also assessed how the required outflow rate compares with an empirical calibration. 
In our sample the molecular gas mass goes down from $\sim10^{10}$ at an age of 1\,Gyr to a few times $10^9\,\msun$ at 9\,Gyr. Hence, the average outflow rate during this time must be around $1\msunyr$.

\begin{figure*}
\includegraphics[width=0.8\textwidth]{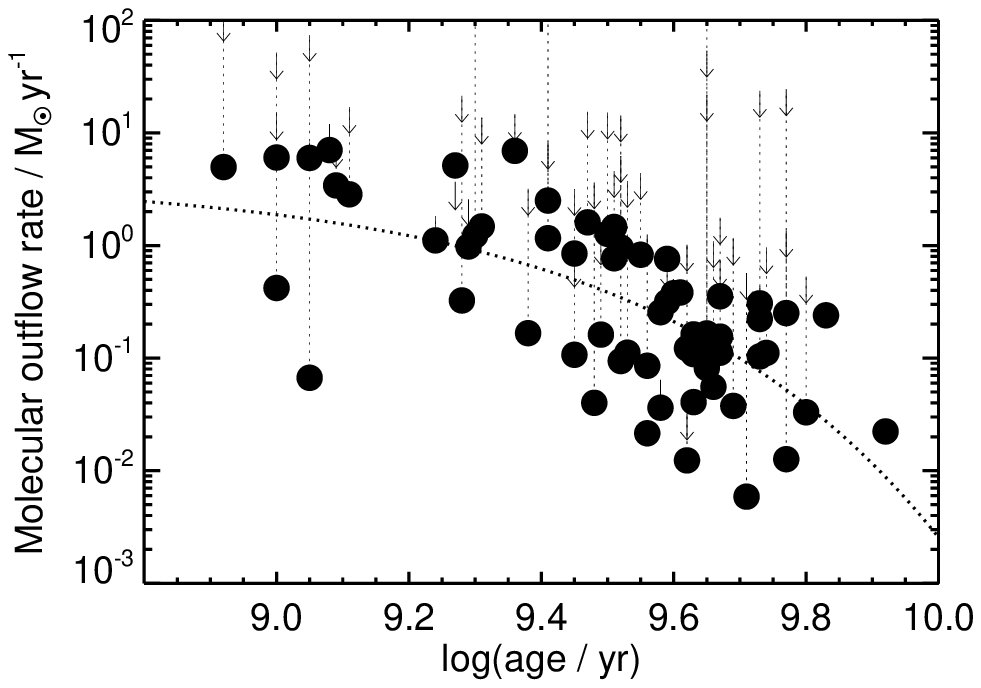}
 \caption{Molecular outflow rate (eq.~5 of \citealt{fluetsch19}) as a function of luminosity-weighted stellar age. {\it Circles} denotes the measurement without taking into account the AGN contribution (because most of the galaxies in our sample have not been detected at {\oiii}). {\it Arrows} denote the molecular outflow rates with the maximum ($2\sigma$) contribution of AGN allowed by the {\oiii} data.
 The dashed line is an exponential fit to the data with the form $(\mbox{outflow rate / \msunyr})=(3.9\pm1.0)\exp\{-{\rm age}/[(1.36\pm   0.24)\,{\rm Gyr}]\}$.
 These outflow rates are too low to explain the ISM decline 
 (see Fig.~\ref{fig:mdms_age_outf_astra}).
}
 \label{fig:out_age}
\end{figure*}

We estimated the actual outflow rate from the empirical calibration of \citet[][their eq.~5]{fluetsch19} based on SFR, $\mstar$ and AGN luminosity ($L_{\rm AGN}$). We estimated the AGN luminosity from the {\oiii} luminosity ($L_{\rm [O III]}$), as done in \citet{fluetsch19} for cases with no X-ray data: $L_{\rm AGN}=3500 L_{\rm [O III]}$ \citep{heckman04}. Only five galaxies in our sample have $>3\sigma$ detection of the {\oiii} line, so we first calculated the molecular outflow rates setting the AGN luminosity to zero. This is shown as circles in Fig.~\ref{fig:out_age}. Then we calculated the maximum outflow rate by setting the AGN luminosity to the $2\sigma$ upper limit allowed by the {\oiii} data (arrows on this figure). It is clear that the measured outflow rate are at the level of the required average rate only at the beginning of the evolution. Hence, the outflows calculated in this way are not powerful enough to explain the gas mass decline throughout the entire period. 

To demonstrate this we fitted an exponential function to the outflow rate as a function of age and obtained the functional form of the outflow rate in {\msunyr} of $(3.9\pm1.0)\exp\{-{\rm age}/[(1.36\pm   0.24)\,{\rm Gyr}]\}$ (the half-life time is $0.41 \pm 0.07$\,Gyr, a dashed line on Fig.~\ref{fig:out_age}). Then starting at the molecular mass of $10^{10}\,\msun$ at an age of 1\,Gyr we show in 
Fig.~\ref{fig:mdms_age_outf_astra} in the Appendix
how much gas and dust is removed with the outflow of such strength. It is clear that the evolution is nearly flat, as the outflow rate is too low (especially at older ages) to explain the gas decline.

\subsection{Dust grain destruction}
\label{sec:graindes}

Dust grains can be destroyed e.g.~by SN shocks, AGN or by sputtering in the hot gas. This scenario predicts the following.

\begin{enumerate}
\item $M_{\rm gas}/\mdust$ should increase with age as gas is not destroyed, just dust. \label{pred:grain:gdr}
\item {\mstar} should be constant or slightly increasing with age, given low SFRs of galaxies in our sample.
\end{enumerate}

This mechanism is ruled out by prediction \ref{pred:grain:gdr}. The gas-to-dust ratio is not increasing with age (Fig.~\ref{fig:mh2md_age}), as directly predicted by this mechanism. The gas masses are also decreasing with time (Fig.~\ref{fig:mdms_age}).

\subsection{Astration/strangulation}
\label{sec:starv}

Gas and dust are incorporated into newly formed stars (astration), so 
 if the SFR is higher than the gas inflow rate, or if the gas inflow is stopped  \citep[a process called strangulation;][]{peng15,trussler20}, then
the gas reservoir will be depleted 
This scenario predicts the following.

\begin{enumerate}
\item $M_{\rm gas}/\mdust$ should be constant with age, as both components are being removed, or slightly increasing if dust distribution is more clumpy and this is where star formation is occurring.  \label{pred:astr:gdr}
\item {\mstar} should be constant or slightly increasing with age, given low SFRs of galaxies in our sample.
\item Given the decreasing SFR with age (Fig.~1 in \citetalias{michalowski19etg}), the dust and gas removal should be weaker at older ages, so the drop of $\mdust/{\mstar}$ and $M_{\rm gas}/{\mstar}$ should flatten at older ages (see the $\mdust$ panel in Fig.~1 in \citetalias{michalowski19etg}). \label{pred:astr:flat}
\end{enumerate}

This mechanism is ruled out by prediction   \ref{pred:astr:flat} and, with a lower statistical significance, by prediction \ref{pred:astr:gdr}.
First, Fig.~\ref{fig:mdms_age} clearly shows steepening of the decline of the $\mdust/{\mstar}$ ratio, so the dust removal is stronger at high ages. Hence this cannot be connected with the SFR, which is the lowest at high ages.
Indeed, \citetalias{michalowski19etg} showed that the effect of astration with measured SFRs on the dust amount is minor, not able to explain the dust decline.

In order to quantify what effect the astration might have on gas evolution we made calculations similar to those presented in \citetalias{michalowski19etg} for dust evolution. Fig.~1 there presents an exponential fit to the SFR evolution (see eq,~\ref{eq:sfr_age} above). We used this, starting with $10^{10}\,\msun$ of gas to see the effect of astration. This is shown in 
Fig.~\ref{fig:mdms_age_outf_astra} in the Appendix.
The evolution is nearly flat, showing that the SFRs are not high enough, so astration cannot be responsible for  most of the gas and dust decline. Hence, we conclude that, while astration is happening, the effect is too weak to fully explain the data. This is consistent with the conclusion drawn in \citetalias{michalowski19etg}.

Similarly, Fig.~\ref{fig:mdms_age_model} in the Appendix shows that models with only astration are inconsistent with the data because the gas removal is the weakest for high ages. Only incorporating additional gas removal makes the models agree with the data.

Finally, the gas-to-dust ratio is not constant (or increasing) with age (Fig.~\ref{fig:mh2md_age}). The Spearman correlation coefficient for that plot is  $-0.85$ with a probability of a null hypothesis (no decline) of $\sim0.0037$ ($\sim3\sigma$).

\subsection{A decline in the number of dust-producing AGB stars}
\label{sec:agbnumber}

It could be that the decline of $\mdust/{\mstar}$ and $\mgas/{\mstar}$ is due to a decline in the number of dust-producing AGB stars at older ages, which also release gas to the ISM. 
This is only possible if the dust and gas produced by these AGB stars dominate the current dust and gas budget in these galaxies. Otherwise, if their contribution is minor, then the decline in their numbers cannot explain the decline in the dust or gas content. 

\citet{rowlands12} and \citetalias{michalowski19etg} showed that indeed AGB stars do not contribute significantly to dust production in these galaxies. Given their stellar masses, the number of AGB stars is too low and each of these stars would need to have produced a too high amount of dust.

We now examine whether the gas masses we detected in the ETGs could have had a large contribution from the ejecta of AGB stars. We compare the gas masses with the expected numbers of AGB stars calculated from the stellar masses, using a similar argument as presented in \citet{michalowski10qso,michalowski10smg4}, \citet{michalowski15}, and \citet{lesniewska19} \citep[see also][]{morgan03,dwek07,dwek11,rowlands14b}.
A stellar mass of $10^{11}\,\msun$ implies around $10^{10}$ AGB stars ($1.5-8\,\msun$), assuming the \citet{chabrier03} IMF. A gas mass of $\sim10^{10}\,\msun$ for the galaxies in our sample at the lowest ages implies that each AGB star would need to release approximately $1\,\msun$ of gas. This is the order of magnitude of the AGB total ejecta mass \citep{morgan03,ferrarotti06}. This calculation neglects past gas consumption due to star formation and outflows (which would increase the required total ejecta mass per star) and inflows (which would decrease the required ejecta mass). It seems that AGB stars are numerous enough to explain the observed gas masses. However, as mentioned above, they do not contribute significantly to dust production, so it is likely that the gas decline is connected with the same mechanism as the dust decline, and not a decreasing number of AGB stars. On the other hand, this calculation supports the internal origin of the ISM in these ETGs, as claimed in \citetalias{michalowski19etg}.

This leaves the question on why the majority of ETGs  selected in a similar way to our sample do not have detectable ISM content, given that they have similar stellar masses, so a similar number of AGB stars. We speculate that this may be due to the non-detected galaxies being older, or exhibiting stronger ISM destruction mechanism (e.g.~more powerful AGN), or having a higher amount of very hot gas, leading to immediate ionization of all the gas released by AGB stars.

\subsection{Dust cooling}
\label{sec:cool}

It could be that with time dust is not destroyed, but cools down, so becomes invisible for {\it Herschel}. This scenario predicts the following.

\begin{enumerate}
\item As a result of this extra cold dust component, longer-wavelength observations should result in high flux levels above the extrapolation from the {\it Herschel} wavelengths.
\item Dust temperatures measured with the {\it Herschel} data should decrease with age.
\item Gas-to-dust ratios should increase, because the gas mass measurement is unaffected. 
\end{enumerate}

The data are inconsistent with all these predictions.
Our JCMT/SCUBA-2 data rule out this mechanism, because we do not see the flux excess at $850\,\micron$  
(Fig.~5 in \citetalias{michalowski19etg}).
Moreover, we do not see any trend of the dust temperature with age and the range of the temperatures is relatively narrow (Fig.~1 in \citetalias{michalowski19etg}). Finally, the gas content also decreases (Fig.~\ref{fig:mdms_age}) and hence, the gas-to-dust ratios are constant or slightly decreasing (Fig.~\ref{fig:mh2md_age}). 

\subsection{Dust heating}
\label{sec:heat}

If dust is progressively heated to high  temperatures (but not destroyed), then it could become invisible for {\it Herschel}/SPIRE.
This scenario predicts the following.

\begin{enumerate}
\item The SEDs of these galaxies should have the peak shifted towards shorter wavelengths.
\item Dust temperature measured with {\it Herschel} data should increase with age.
\item Gas-to-dust ratios should increase, because the gas mass measurement is unaffected. 
\end{enumerate}

The data are inconsistent with all these predictions.
The SEDs of the galaxies in our sample do not show any evidence of a  shift of the peak towards shorter wavelengths (
\citealt{rowlands12}). Moreover, we do not see any trend of the dust temperature with age (Fig.~1 in \citetalias{michalowski19etg}).
Finally, the gas-to-dust ratios do not increase with age (Fig.~\ref{fig:mh2md_age}).

An additional warm component with a temperature of $30$\,K could only contribute a few percent to the cold dust mass in order not to overshoot the $100\,\micron$ detections or limits. The contribution of even hotter dust could be even lower because even a small amount of hot dust is too bright. Hence, the possible contribution of hot dust is too small to explain the decrease of the $\mdust/{\mstar}$ ratio.

\subsection{Environmental influence}
\label{sec:env}

In principle it is possible that the observed trend is the reflection of environmental influence. Galaxies in richer environments are quenched quicker (so the stellar ages are higher) and the ISM is being removed quicker due to environmental effects like interactions or ram-pressure stripping.

This scenario predicts the following.

\begin{enumerate}
\item There should be a trend of the environmental density with age (and therefore with the $\mdust/{\mstar}$ ratio).
\item These galaxies should live in rich environments in which the influence on their ISM is significant. 
\end{enumerate}

None of these predictions are consistent with the data. As shown in Fig.~1 in \citetalias{michalowski19etg}, the galaxy density does not depend on the age.
Moreover, these galaxies do not reside in rich environments. The  projected galaxy densities are below 10\,Mpc$^{-2}$, below those of galaxy groups. We also do not see any dependence of the dust decline on environment in the extended analysis of 2\,000 dusty ETGs \citep{lesniewska23}.

\subsection{Mergers with gas-rich galaxies}
\label{sec:merger}

It could be that the trend cannot be interpreted as a time evolution, but results from mergers of passive gas-poor galaxies with gas-rich star forming galaxies of different masses, or that each passive galaxy experienced different number of mergers. If the merging star-forming galaxy was relatively large (or a passive galaxy had experienced more mergers), then the resulting mean age would be low (because of many young stars brought in during mergers) and the resulting $\mdust/{\mstar}$ would be high (because more dust is brought). A merger with a small star-forming galaxy (or fewer number of mergers) would result in a much higher derived age and a much lower $\mdust/{\mstar}$ ratio. This scenario predicts the following.

\begin{enumerate}
\item $M_{\rm gas}/\mdust$ should be increasing with the derived age, because passive galaxies with high derived ages would need to have merged with smaller galaxies, which have high $M_{\rm gas}/\mdust$ ratios \citep{grossi10, grossi15,galametz11,leroy11, cortese12,hunt14b,remyruyer14}. \label{pred:merg:gdr}
\item {\mstar} should be constant or slightly decrease with the derived age because merging dwarf galaxies would not contribute much to {\mstar}, and the
galaxies in our sample with lower inferred ages would have on average experienced more merging and so may be slightly more massive.\label{pred:merg:mstar}
\item $M_{\rm gas}/{\mstar}$ should be decreasing with the derived age, for the same reason why $\mdust/{\mstar}$ is decreasing (higher-mass galaxies bring more gas).
\item $\mdust$ should only be weakly correlated with {\mstar}, because the gas-rich companions would not bring significant amount of stars, so the final {\mstar} should only weakly depend on the number of merger events.
\label{pred:merg:mduststar}
\item Sizes should be increasing with decreasing age, because in this scenario a low derived age means more merging and galaxies grow in size as a result of mergers \citep{naab09,hopkins10b,trujillo11,furlong17}.
\label{pred:merg:size} 
\end{enumerate}

This mechanism is ruled out by predictions \ref{pred:merg:gdr}, \ref{pred:merg:mstar}, \ref{pred:merg:mduststar}, and \ref{pred:merg:size}. It has also been ruled out by \citet{rowlands12} and \citetalias{michalowski19etg} in the context of the analysis of the source of dust in these galaxies \citep[see also][]{donevski23}.

The gas-to-dust ratio is not increasing with age (Fig.~\ref{fig:mh2md_age}), inconsistent with prediction \ref{pred:merg:gdr}.  
Fig.~1 in \citetalias{michalowski19etg}
shows that the stellar mass is increasing slightly with age, which should not be the case if the apparent dust decline with age was a result of merging with smaller (or fewer) galaxies at high ages and larger (or more numerous) galaxies at low ages (prediction \ref{pred:merg:mstar}). Moreover,  Fig~3 in \citetalias{michalowski19etg} shows that {\mdust} is correlated with {\mstar}, 
inconsistent with prediction \ref{pred:merg:mduststar}. The Spearman rank correlation coefficient is 0.5 with a very small probability ($\sim3\times10^{-5}$, $\sim4\sigma$) of the null hypothesis (no correlation) being acceptable. 
Fig.~2 in \citetalias{michalowski19etg}
also shows that there is no correlation of the size of these galaxies with age, as would be expected if mergers are the mechanism responsible for the trends we observe (prediction \ref{pred:merg:size}).

\begin{figure*}
\includegraphics[width=0.8\textwidth]{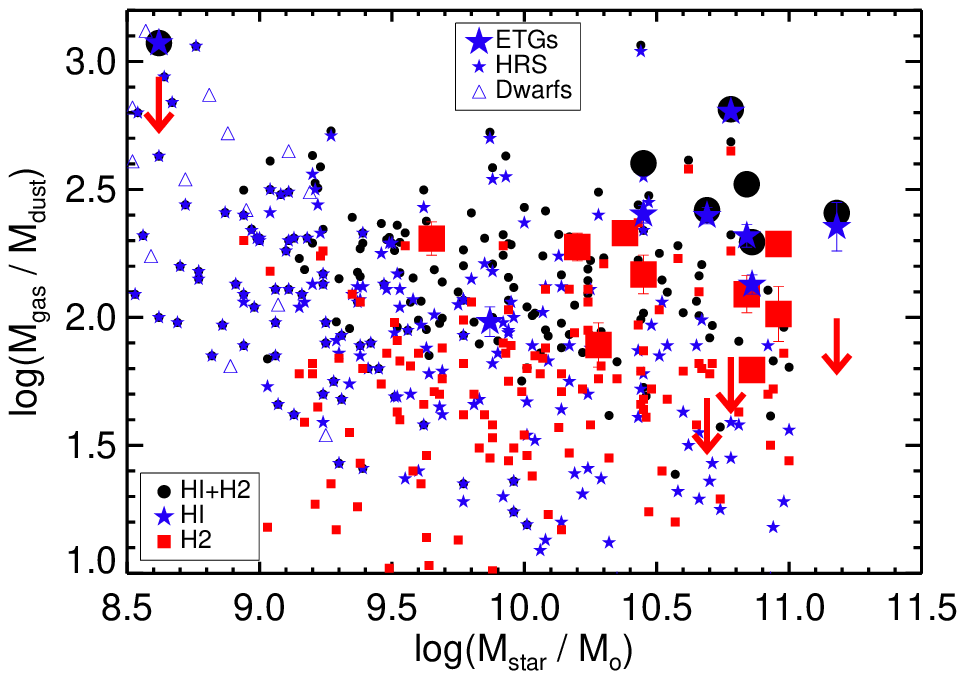}
 \caption{Gas-to-dust ratios as a function of stellar mass. {\it Large symbols} denotes galaxies in our ETG sample, and {\it small symbols} denote those from HRS. 
 {\it Red squares}, {\it blue stars}, and {\it black circles} denote molecular gas, atomic gas, and total gas, respectively (the latter only for galaxies with both CO and {\hi} measurements).
 ETGs in our sample have high gas-to-dust ratios, but consistent with the HRS population arguing against external source of the ISM by merging with dwarf galaxies, which have high gas-to-dust ratios.
}
 \label{fig:mh2md_ms}
\end{figure*}

Conversely, \citet{davis15} concluded that the gas in most of ETGs with dust lanes is of external origin, because of the large range of gas-to-dust ratio extending to high values ($\sim800$), typical for dwarf galaxies, which have apparently merged with gas-poor ETGs \citep[see also][]{lianou16}. 
 For our sample the gas-to-dust ratios are smaller, 
 suggesting high-metallicity ISM, and hence internal origin. We show the gas-to-dust ratio as a function of stellar mass in Fig.~\ref{fig:mh2md_ms}. We compare our galaxies with spirals from the {\it Herschel} Reference Survey \citep{boselli10,boselli14,bendo12,cortese12b,cortese14,ciesla12,ciesla14}. Their dust is likely of internal origin (because they are currently star-forming) and the ETGs in our sample fully overlap with these spirals, which supports the internal origin of dust in these ETGs. On the other hand, dwarf galaxies exhibit much higher atomic gas-to-dust ratios of 300--10\,000 \citep{grossi10,grossi16,cormier14,hunt14b,hunt15,hunt17}.

\subsection{Stellar mass-age correlation}
\label{sec:magecorr}

More massive galaxies are on average older
\citep[e.g.][]{mcdermid15}, so in principle the $\mdust/\mstar$-age anti-correlation could be driven by the $\mstar$-age correlation. This scenario predicts that this anti-correlation should disappear if only a narrow range of stellar masses is analysed. 

The sample analysed in this paper is too small to subdivide it in narrow bins of stellar mass, but this can be done for galaxies selected similarly by \citet{lesniewska23} in a much bigger field. Fig.~\ref{fig:mdms_age_narrowms} in the appendix shows that the $\mdust/\mstar$-age anti-correlation persists even for very narrow ranges of stellar masses, which is inconsistent with it being driven by the $\mstar$-age correlation. The scatter in the stellar mass bins $\log(\mstar/\msun)=10.5$--$10.6$ and $10.6$--$10.7$ is 0.39 and 0.36\,dex, respectively, so it does not increase compared to the full sample with a scatter of 0.43\,dex.

\subsection{Selection bias}
\label{sec:bias}

Our sample has been selected based on redshift, elliptical morphology and dust detection and is a $250\,\micron$ flux-limited sample. If the $\mdust/\mstar$-age decline was the a result of a selection bias, then this would need to imply that
\begin{enumerate}
    \item The selection criteria should remove old very dust-rich galaxies. 
    \item The selection criteria should remove young galaxies with dust content similar to that detected in older galaxies.
\end{enumerate}
None of these biases are introduced by our selection criteria. If old galaxies with high dust content or young galaxies with dust content equally low as we see for older galaxies existed, we would have detected them.

\section{Energy source}
\label{sec:esource}

While observations favor a scenario involving
the removal of the entire cold ISM or outflows, the question remains what is
the source of energy responsible for the ISM removal. This cannot be connected with the current star formation, because the SFRs are low in the galaxies in our sample and the SFR level gets progressively lower, which would result in a flattening evolution of the ISM content, inconsistent with the observations. 
We consider planetary nebulae (PNe), SNe Type Ia, cosmic rays, hot halo gas, and AGNs. PNe or SNe Type Ia (either directly or through cosmic rays), or AGNs (if their duty cycle is low) are the most likely explanations.
They are  short lived sources of energy, so consecutive generations of stars need to constantly go through these phases to make these scenarios possible.

For all potential energy sources, we will discuss how the state  of the cold gas changes. It can either be removed from the galaxy entirely, transformed into the warm ionized gas (temperature of $10^{4-5}$\,K) or hot ionized gas (temperature $>10^6$\,K). Given the lack of X-ray data, we do not have a constrain on the latter, but we calculated the amount of warm ionized gas from the H$\alpha$ luminosity using eq.~1 of \citet{pagotto21}, following their assumption of electron density of $n_e=100\,\mbox{cm}^{-3}$. The resulting masses of warm ionised gas of $10^{4-5}\,\msun$ are much lower than the difference between the cold gas masses of the youngest and oldest galaxies analysed here of $10^9\,\msun$. Hence, cold gas cannot be transformed into the warm ionised phase for these galaxies and can only be heated to much higher temperatures or removed by outflows.   

\subsection{Planetary nebulae}

Low-mass stars during the main-sequence or AGB phases are not energetic enough to ionize the gas around them (which is required to explain the decline of both molecular and atomic gas). However, the post-AGB/PN phase is a possibility, usually referred to as a hot low-mass evolved star \citep[HOLMES;][]{binette94,floresfajardo11,herpich18}. 
 During this phase a star ejects its envelope with an expansion velocity of around $30\,\kms$, and this gas has the temperature of $\sim10^{4-5}$\,K \citep[e.g.][]{cuisinier96,milingo02,sharpee07,sahai10,bohigas15,ali15,ali19}. 
This is enough to move gas to the warm ionized phase, but is not enough to expel gas from a galaxy or to heat it to very high temperatures of $>10^6$\,K. However, in environments with high  stellar velocity dispersion, the interaction of PNe with the ambient gas can lead to heating to such high temperatures \citep{conroy15}.
This scenario was invoked by \citet{conroy15} as a mechanism responsible for preventing star formation in quiescent galaxies.
There are indeed indications that post-AGB stars are responsible for photoionization of gas in ETGs \citep{binette94,stasinska08,cidfernandes10,cidfernandes11,sarzi10,kehrig12,yan12,papaderos13,singh13,gomes16,herpich18}.

We made simple calculations to estimate whether PNe are numerous enough in the galaxies in our sample to explain the gas decline. A galaxy with a total stellar mass of $\sim10^{11}\,\msun$ has $\sim1.9\times10^{10}$ stars with masses between $1$--$8\,\msun$, which ended their lives as PNe. In our sample the total gas mass goes down from $\sim10^{10}$ at the age of 1\,Gyr to $\sim2\times10^9\,\msun$ at 9\,Gyr. Hence, one PN would need to remove $\sim8\times10^{9}\msun / 1.9\times10^{10}=0.4\msun$ of gas.


This calculations does not take into account that some stars go through the PN phase before the period considered here for gas removal of 8\,Gyr. At $z=0.1$-$0.3$ there are additional 2-4\,Gyr after the Big Bang. If we make the most conservative assumption that all stars were created at the time of the Big Bang, then during this extra time stars with masses of more than $1.9$--$1.4\,\msun$ have already went through the PN phase before our considered ISM-removal phase. This decreases the number of available PNe to $1.1$--$0.8\times10^{10}$ and therefore increases the required gas removal to $0.7$--$1\,\msun$ per PN.

The largest (i.e.~oldest) PNe or circumstellar envelopes of evolved stars  have radii of the order of 1\,pc \citep[e.g.][]{odell04,sahai10,matthews15}. For an ISM density of $20$--$50\,\mbox{cm}^{-3}$ for the cold atomic medium \citep{ferriere01}, this corresponds to $\sim2$--$5\,\msun$ of swept-up gas.
This is similar to our estimate of the required gas removal per PN, so they are numerous enough to be responsible for the detected gas decline.

\subsection{Supernovae Type Ia}


  

SNe Ia can ionize gas and destroy dust in the swept up part of the ISM \citep{li20}. The swept-up gas can reach temperatures of $>10^6$\,K \citep{ceverino09,hopkins18,li20}, required to move it to the very hot gas phase.

Similarly as for PNe we estimate the required effectiveness of SN Ia. Using eq.~19 and Table~1 of \citet{andersen18} we estimate the rate of SN Ia for a galaxy with $\mstar=10^{11}\,\msun$ and $\mbox{SFR}=0$--$10\,\msunyr$ to be $0.008$--$0.014\,\mbox{yr}^{-1}$. For a period of 8\,Gyr this corresponds to $\sim1.1$--$6.4\times10^7$ SNe Ia. Hence, one SN Ia would need to remove $\sim730$--$130\,\msun$ of gas.


SNe Ia release around $10^{51}\,$erg of energy  \citep{khokhlov93}. For a SN with such energy in the simulations of \citet{yepes97} $\sim20$--$5000\,\msun$ of gas is ionized. 
The required strength of SN feedback for our sample is consistent with this range.
 We therefore find this mechanism feasible.

We do not consider core-collapse SNe here, because the lifetimes of their progenitors are very short, so their numbers are closely connected with recent star formation. Therefore their numbers decrease for galaxies with high ages, which makes them unable to explain the detected gas decline, similarly to astration.
  
\subsection{Cosmic rays}

The interaction with cosmic rays can lead to the destruction of dust particles by thermal evaporation and sputtering \citep{dwek92}, and also to the heating and ionization of gas \citep{hayakawa61,spitzer68,ferriere01,indriolo09,padovani20,gabici22}. 
The lifetime of cosmic rays is only $10^{6-7}$\,yr \citep{jokipii69,garciamunoz75,garciamunoz77,jokipii76,yanasak01,bisbas15,bisbas17,gabici22}. Hence, cosmic rays produced by core-collapse SNe are unlikely to affect the ISM of the galaxies in our sample, due to their low SFRs, resulting in a low number of currently exploding core-collapse SNe unable to ionize the amount of gas we detect. However, cosmic rays can also be produced by SNe Type Ia \citep{chan19}.

Simulations show that cosmic rays are capable of dispersing or launching gas clouds on the timescale of the order of 10\,Myr \citep{bruggen20}. This is much shorter than the timescale over which the gas content declines  in our sample.
Therefore, in order for this mechanism to be viable, large amounts of cold dust would need to be shielded in dense clouds against the cosmic ray influence, which would make the timescale much longer and consistent with our measurements.

It has been shown with simulations that cosmic rays can reduce SFRs of galaxies by a factor of five and the density of gas in galaxy centers by an order of magnitude \citep{hopkins21,byrne24}.
However, they do not heat gas to temperatures beyond $10^6$\,K, required here. Hence, the only viable mechanism to explain the decline in cold gas reservoir detected in our sample is the CR-driven outflow.

Summarizing, cosmic rays produced by SN Type Ia can also be responsible for the ISM decline we detect if they drive outflows.

\subsection{Active galactic nuclei}

AGN feedback has been invoked as a mechanism of quenching for massive galaxies and they can heat gas to very high temperatures \citep[see the review of][]{fabian12}.

As stated in Section~\ref{sec:outflow}, only five galaxies in our sample have $>3\sigma$ detection of the {\oiii} line, which is used as an AGN indicator. The median $2\sigma$ upper limit on the AGN luminosity, calculated as  $L_{\rm AGN}=3500 L_{\rm [O III]}$ \citep{heckman04} is $2\times10^{42}\,\mbox{erg}\,\mbox{s}^{-1}$. This indicates a low level of current AGN activity. 
This is consistent with our analysis of emission lines of 2\,000 dusty ETGs, where we found that only up to 15\% exhibit  line ratios typical for AGNs (Ryzhov et al.~in prep.).

However, the timescale during which a typical AGN is active is much shorter than the Gyr timescale considered here \citep{novak11,hickox14,padovani17}. Therefore, the visibility of AGN in a given population depends mainly on the duty cycle (fraction of time during which the AGN is active). For a conservatively low Eddington ratio of $0.01$, simulations predict a duty cycle of $\sim1$--10\% \citep[][their Fig.~8]{novak11}, so in a sample fully dominated by AGNs, only this fraction is expected to show current AGN activity. For a higher Eddington ratios (stronger AGN activity) the expected detected fraction is even lower. These fractions are similar to what we obtain, so the AGN feedback can be a possible mechanism of removing the ISM in these galaxies.

\subsection{Hot halo gas}

Hot, X-ray-emitting gas is frequently found in elliptical galaxies with stellar masses similar to those in our sample \citep[e.g.][]{sarzi13,su15,kim19,kokusho19}. 
The dust destruction timescales in hot (10\,000\,K or more) medium are estimated to be $10^4-10^8$ years 
\citep{draine79,jones94,jones04,micelotta10,bocchio12,hirashita15,hirashita17b}. This is much shorter than the dust and gas removal timescale measured here, in \citetalias{michalowski19etg}, and in \citet{lesniewska23}.
However, if the cold ISM is partially shielded from the influence of hot gas, then the destruction process would be slower and could be responsible for the decline we detect.

\citet{smercina18} interpreted the dust decline for post-starburst galaxies as the effect of grain sputtering in a hot medium. However, they did not detect a gas decline, in contrast to the galaxies in our sample. \citet{galliano21} interpreted the low dust-to-gas ratios of ETGs as the result of dust grain sputtering in hot gas, but our sample exhibits much higher dust-to-gas ratios (compare  Fig.~\ref{fig:mh2md_age} and their Fig.~8).

Finally, dust destruction in hot halo gas applies only to low-density medium \citep{bocchio12}. Hence, it is unlikely to be the main mechanism in the galaxies in our sample, given their substantial ISM masses.

\section{Conclusions}
\label{sec:conclusion}
 
We present carbon monoxide (CO) and 21 cm hydrogen ({\hi}) line observations of dusty ETGs and measure the removal of the cold interstellar medium (ISM). We find that all the cold ISM components (dust, molecular and atomic gas) decline at similar rates. This allows us to rule out a wide range of potential ISM removal mechanisms (including starburst-driven outflows, astration, a decline in the number of asymptotic giant branch stars), and artificial effects like stellar mass-age correlation, environmental influence, mergers, and selection bias, leaving ionization by evolved low-mass stars or ionization/outflows by supernovae Type Ia or active galactic nuclei as viable mechanisms. We also provide the support of internal origin of the detected ISM. Moreover, we find that the quenching of star formation in these galaxies cannot be 
explained by a reduction in gas amount alone,
because the star formation rates decrease faster (at a time scale of about 1.8 Gyr) than the amount of cold gas (a timescale of 2.3\,Gyr). Furthermore, the star formation efficiencyies of the ETGs ($\mbox{SFE}\equiv\mbox{SFR}/\mhtwo$) are lower than those of star-forming galaxies, whereas their gas mass fractions ($\fhtwo\equiv \mhtwo/\mstar$) are normal. This may be explained by the stabilization of gas against fragmentation, for example due to morphological quenching, turbulence, or magnetic fields.

\begin{acknowledgements}

We thank Joanna Baradziej and our anonymous referee
for help with improving this paper, 
Claudia Marka for excellent support with IRAM30m observations and reduction, 
 the IRAM30m 
 staff and observers for executing our programs.

 M.J.M., A.L., J.N., O.R., and M.S.~acknowledge the support of 
the National Science Centre, Poland through the SONATA BIS grant 2018/30/E/ST9/00208.
M.J.M.~acknowledges the support of
the Polish National Agency for Academic Exchange Bekker grant BPN/BEK/2022/1/00110,
the National Science Centre, Poland through the POLONEZ grant 2015/19/P/ST9/04010, 
the Polish-U.S. Fulbright Commission,
 UK Science and Technology Facilities Council,
British Council Researcher Links Travel Grant, Royal Society of Edinburgh  International Exchange Programme  and the hospitality of the Instituto Nacional de Astrof\'{i}sica, \'{O}ptica y Electr\'{o}nica (INAOE), the Academia Sinica Institute of Astronomy and Astrophysics (ASIAA), University of Edinburgh, and the Dark Cosmology Centre.
This project has received funding from the European Union's Horizon 2020 research and innovation programme under the Marie Sk{\l}odowska-Curie grant agreement No. 665778.
C.G.~acknowledges funding by the Carlsberg Foundation and the Carnegie Trust Research Incentive Grant (PI: M.~Micha{\l}owski) as well as support through a VILLUM FONDEN Young Investigator Grant (project number 25501).
J.H. was supported by a VILLUM FONDEN Investigator grant (project number 16599).
A.-L.T.~acknowledges the support of 
the National Science Centre, Poland through the POLONEZ grant 2015/19/P/ST9/04010.
T.T.T. has been supported by the Grant-in-Aid for Scientific Research (No. 17H01110 and 21H01128). 
P.B. acknowledges funding through the Spanish Government
retraining plan 'María Zambrano 2021-2023' at the University of Alicante
(ZAMBRANO22-04)
This research was funded in whole or in part by National Science Centre, Poland (grant numbers: 2021/41/N/ST9/02662 and 2020/39/D/ST9/03078).
For the purpose of Open Access, the author has applied a CC-BY public copyright licence to any Author Accepted Manuscript (AAM) version arising from this submission.
This article has been supported by the Polish National Agency for Academic Exchange under Grant No. PPI/APM/2018/1/00036/U/001. 
O.R.~acknowledges the support of 
the National Science Centre, Poland through the grant 2022/01/4/ST9/00037.

This work is based on observations carried out under project number 198-14, 62-15, and 174-15 with the IRAM 30m telescope. IRAM is supported by INSU/CNRS (France), MPG (Germany) and IGN (Spain). 
The research leading to these results has received funding from the European Commission Seventh Framework Programme (FP/2007-2013) under grant agreement No 283393 (RadioNet3). 
The Green Bank Observatory is a facility of the National Science Foundation operated under cooperative agreement by Associated Universities, Inc. 
This research used the facilities of the Canadian Astronomy Data Centre operated by the National Research Council of Canada with the support of the Canadian Space Agency. 

This research has made use of 
the Tool for OPerations on Catalogues And Tables \citep[TOPCAT;][]{topcat}: \url{www.starlink.ac.uk/topcat/ };
SAOImage DS9, developed by Smithsonian Astrophysical Observatory \citep{ds9};
the NASA's Astrophysics Data System Bibliographic Services
and the WebPlotDigitizer\footnote{\url{https://automeris.io/WebPlotDigitizer}} of Ankit Rohatgi.

\end{acknowledgements}


\input{ms2.bbl}


\appendix

\section{Additional figures and tables}

We present here additional figures showing the effect of outflows and astration,  CO and {\hi} spectra, and the effect of the mass-age correlation, as well as a table with model parameters.

%


\newlength{\panelwidth}
\setlength{\panelwidth}{0.3\textwidth}

\begin{figure*}
\begin{tabular}{ccc}
\includegraphics[width=\panelwidth]{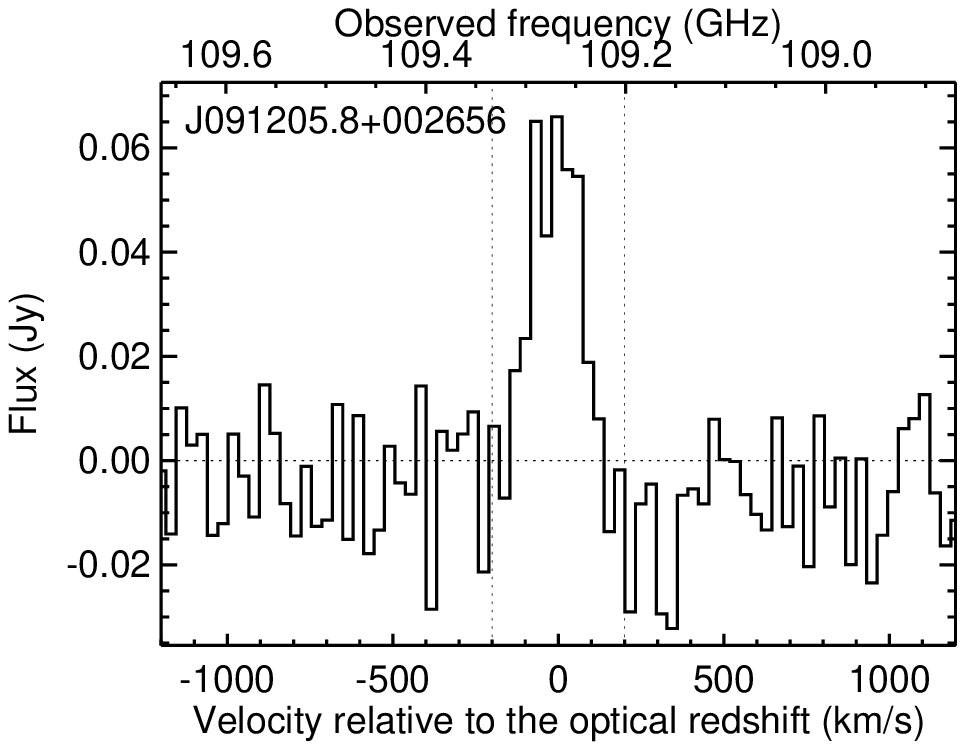}     &  \includegraphics[width=\panelwidth]{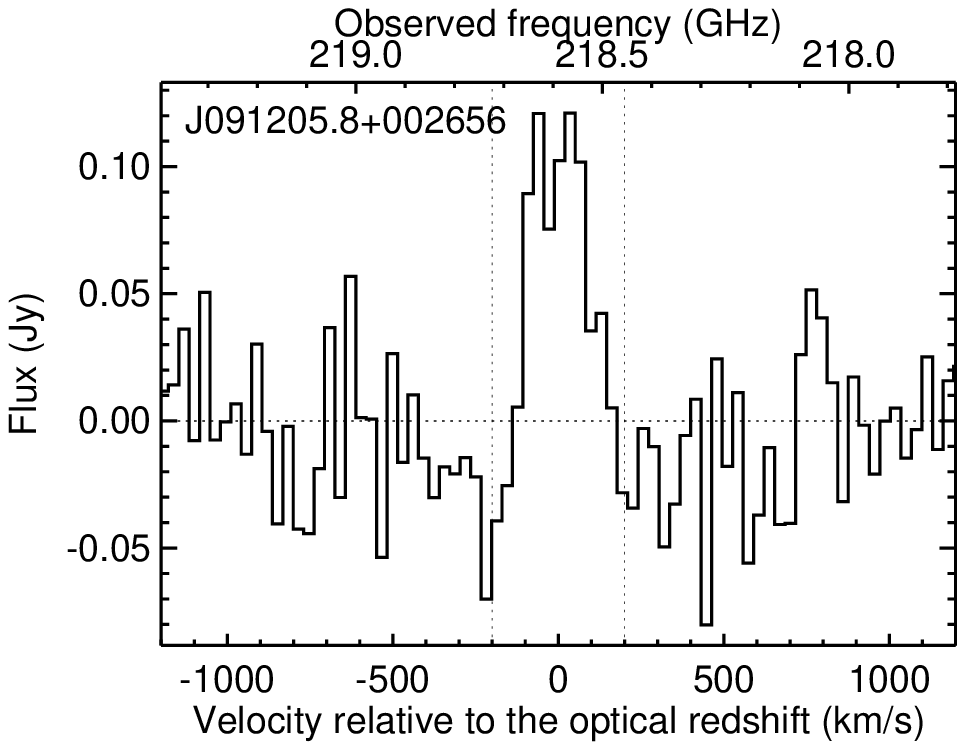} &
\\
\includegraphics[width=\panelwidth]{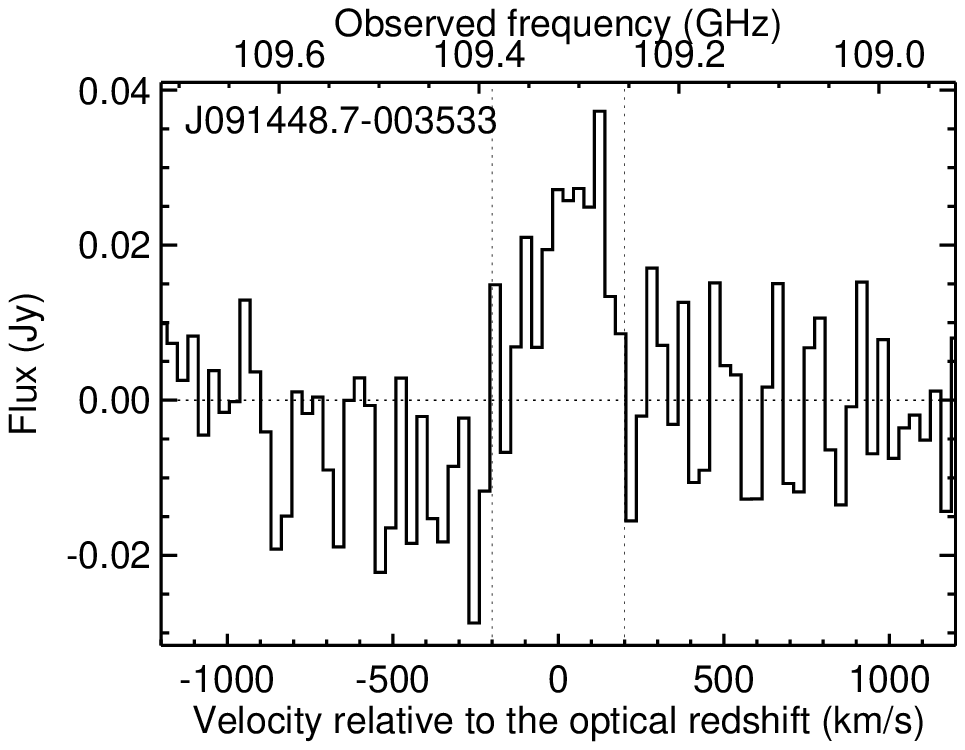}     &  \includegraphics[width=\panelwidth]{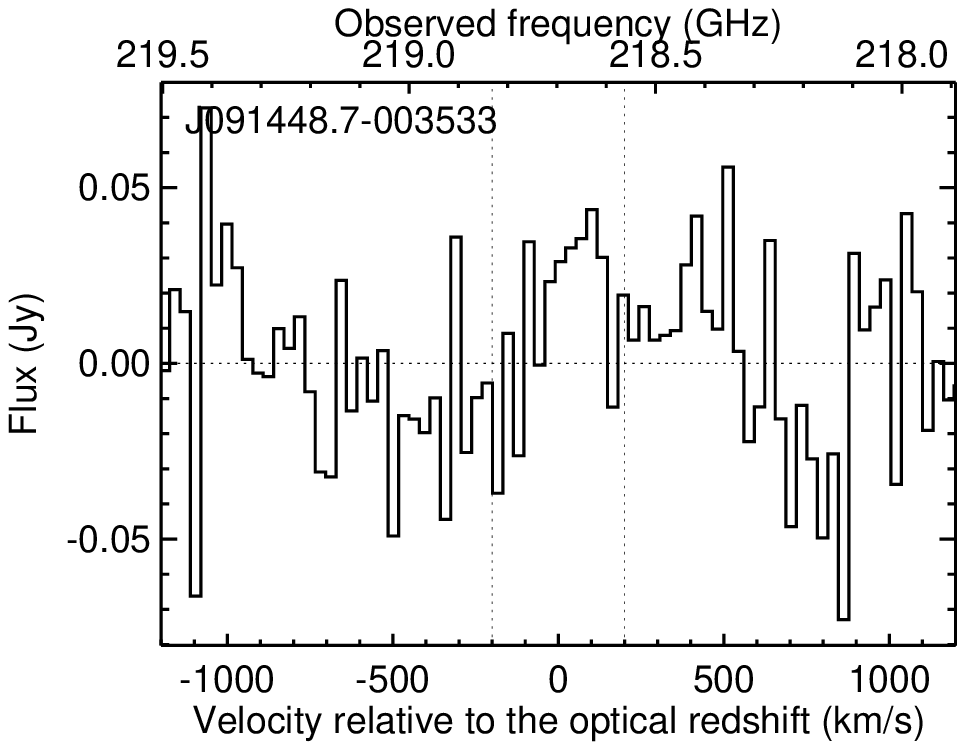} & 
\includegraphics[width=\panelwidth]{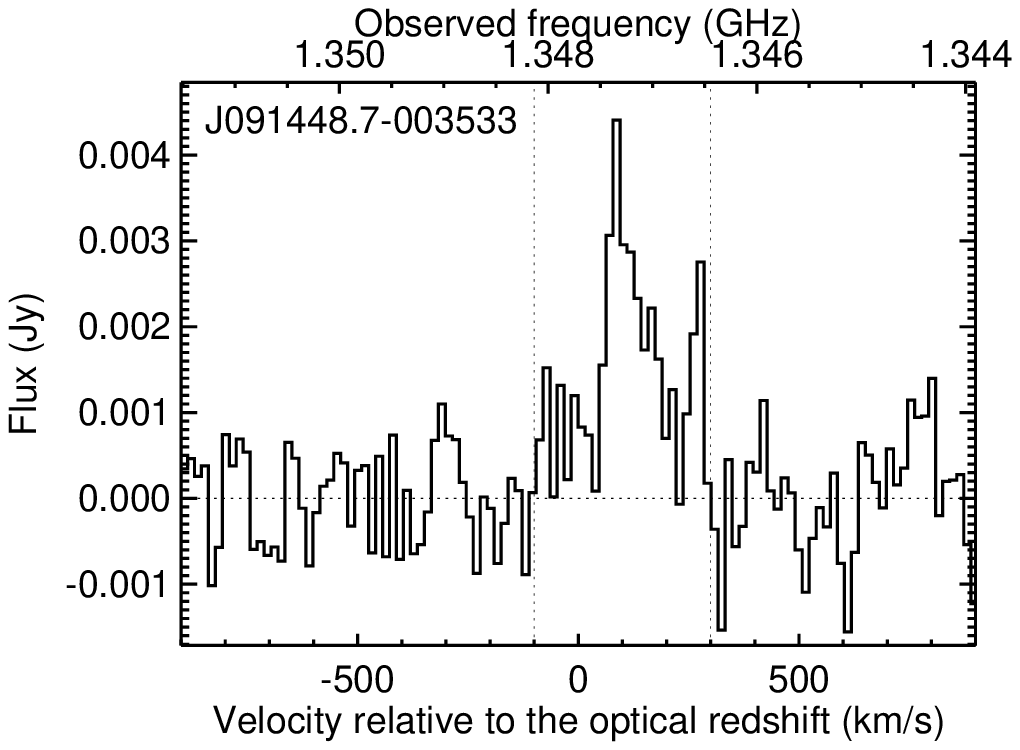}
\\
\includegraphics[width=\panelwidth]{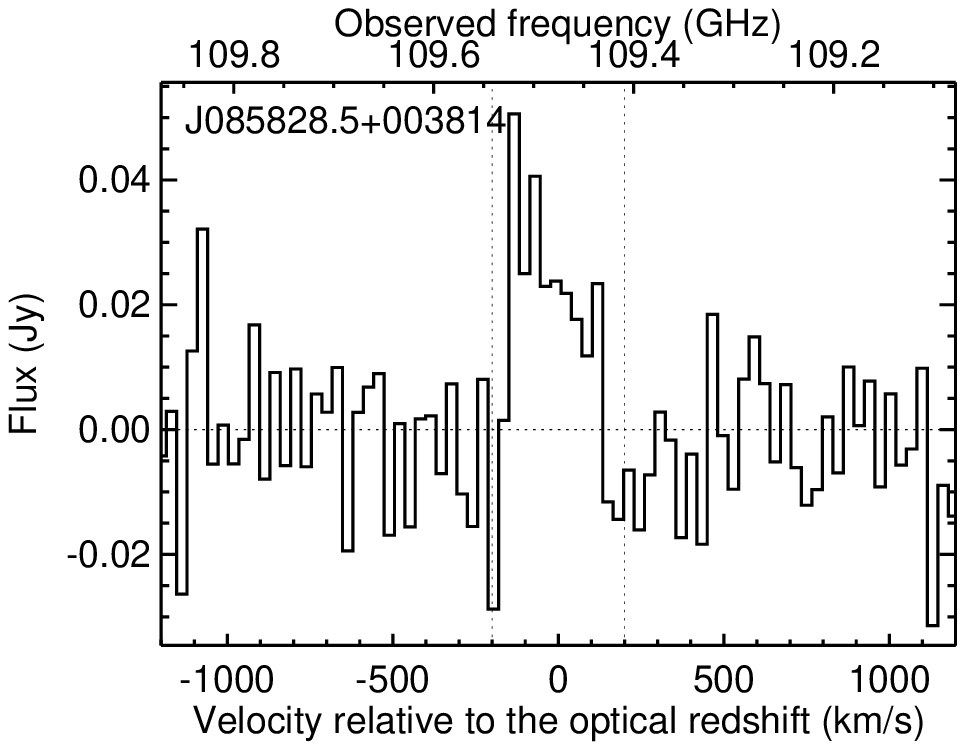}     &  \includegraphics[width=\panelwidth]{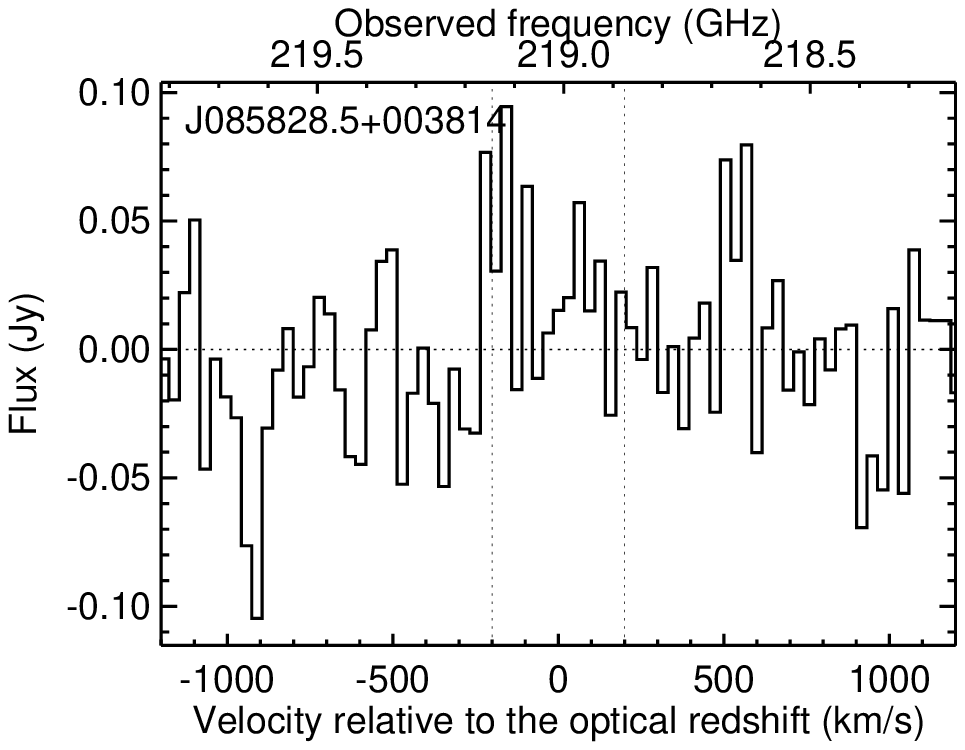} & 
\includegraphics[width=\panelwidth]{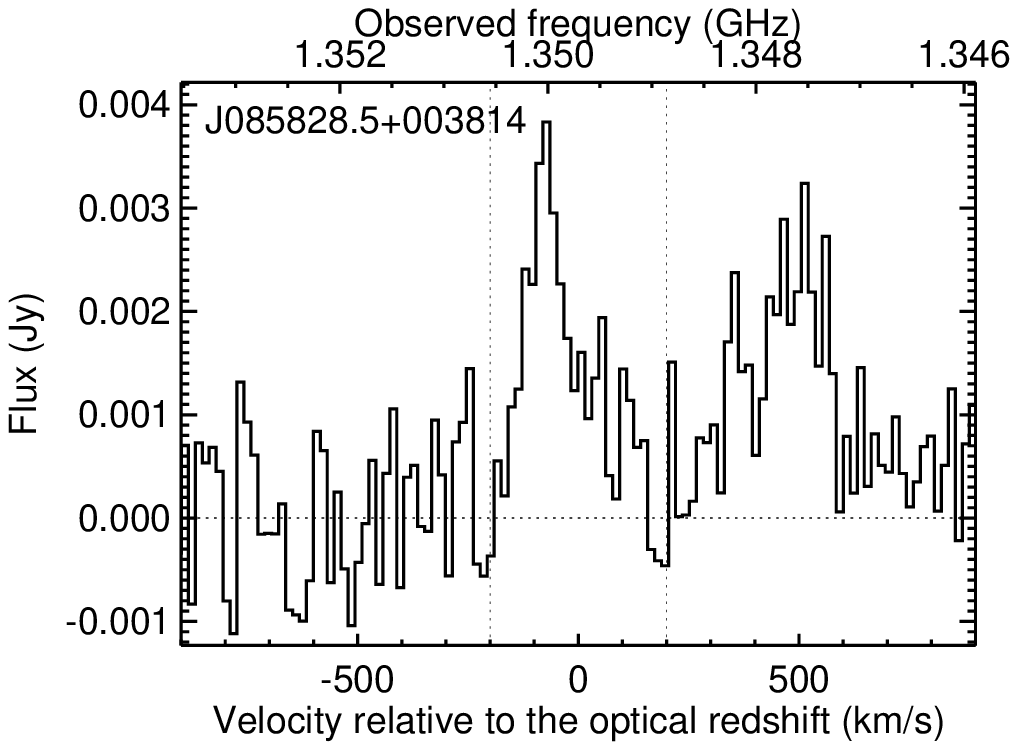} \\
\includegraphics[width=\panelwidth]{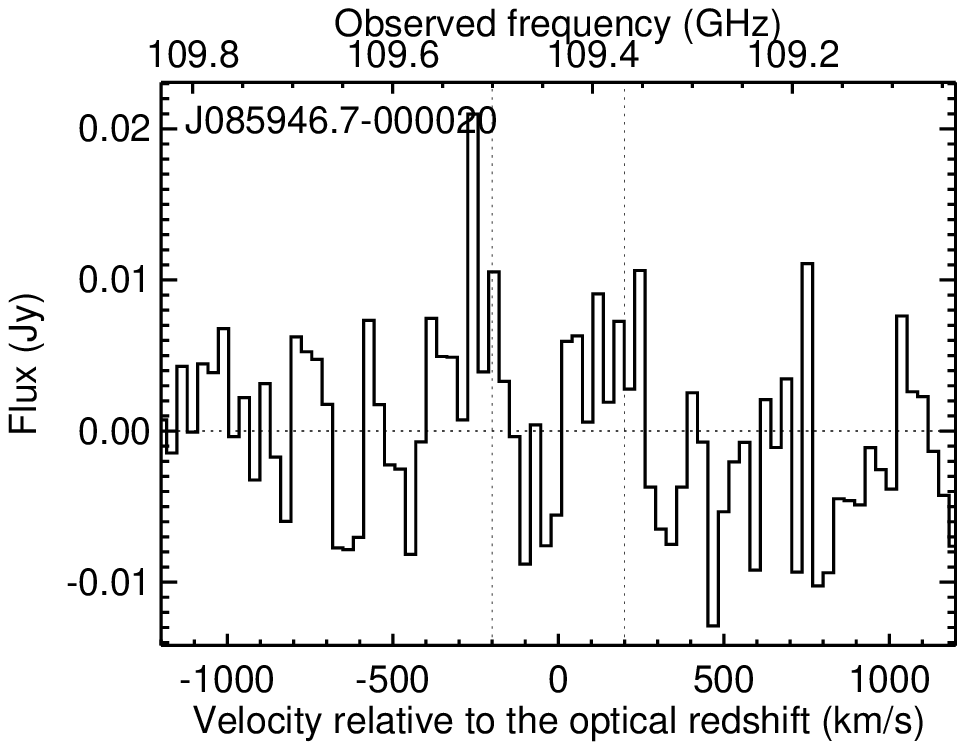}     &  \includegraphics[width=\panelwidth]{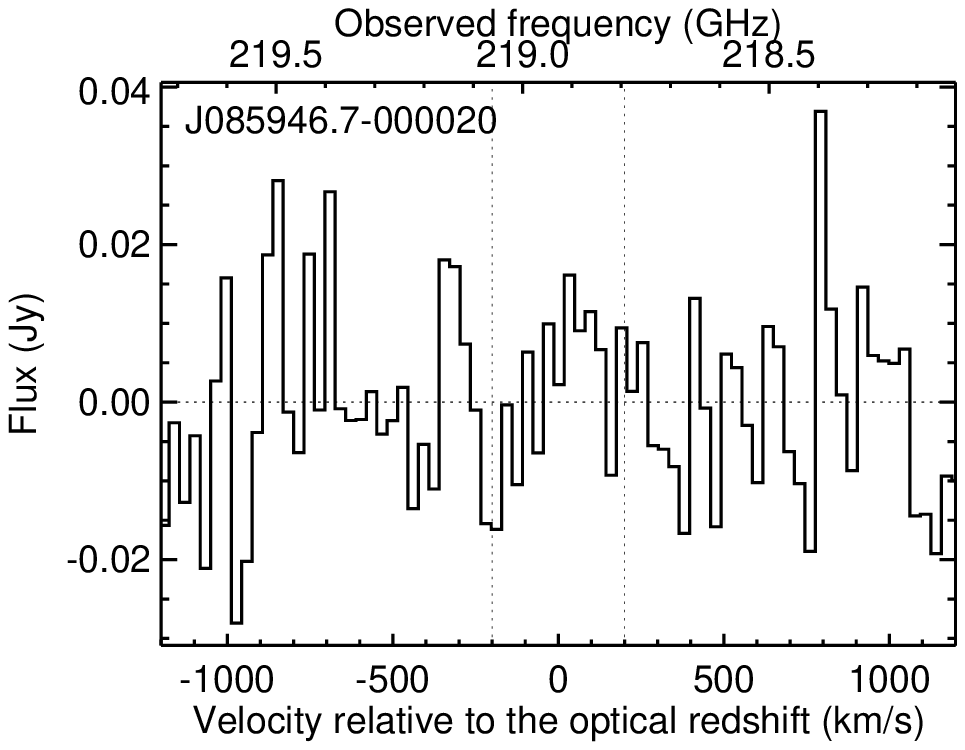} & 
\includegraphics[width=\panelwidth]{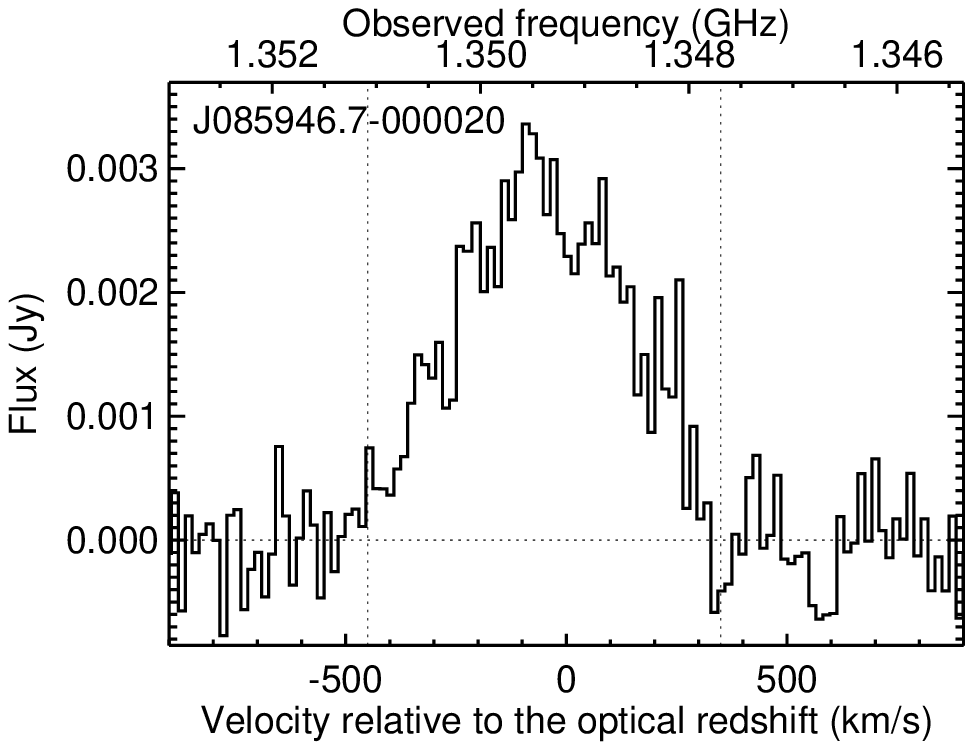} \\
\includegraphics[width=\panelwidth]{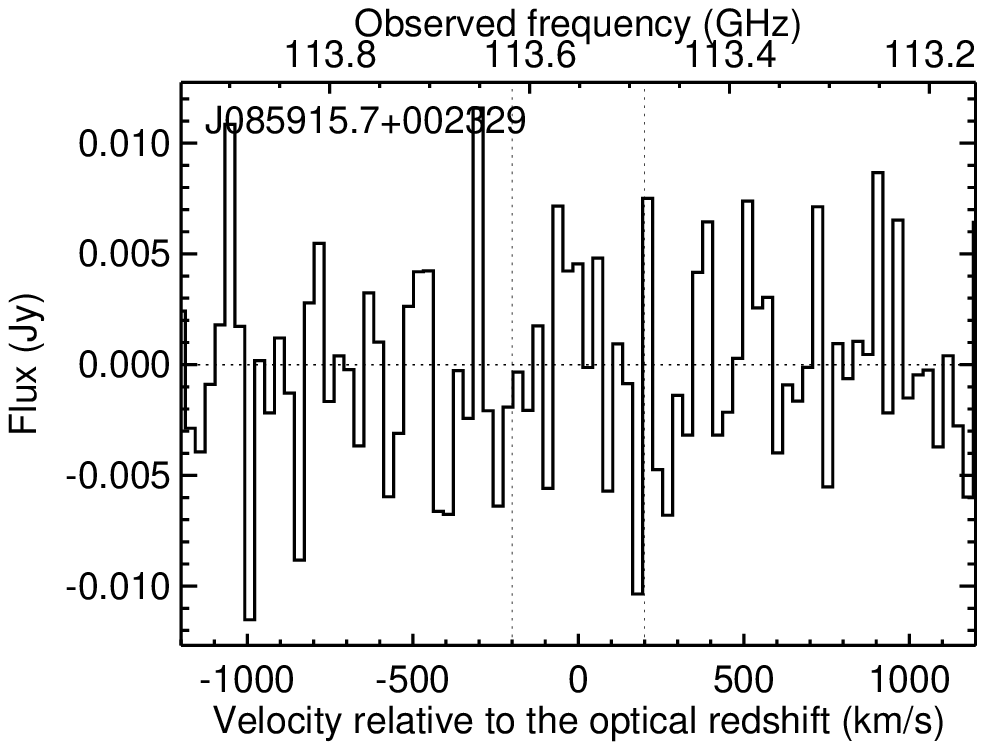}  & & 
\includegraphics[width=\panelwidth]{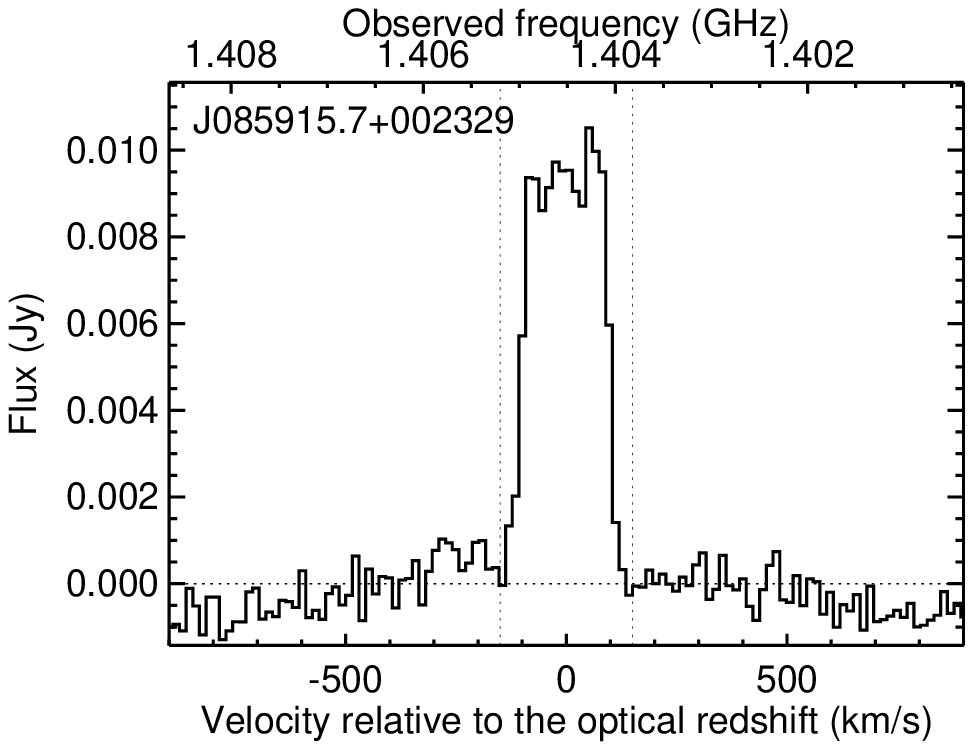}    \\
\includegraphics[width=\panelwidth]{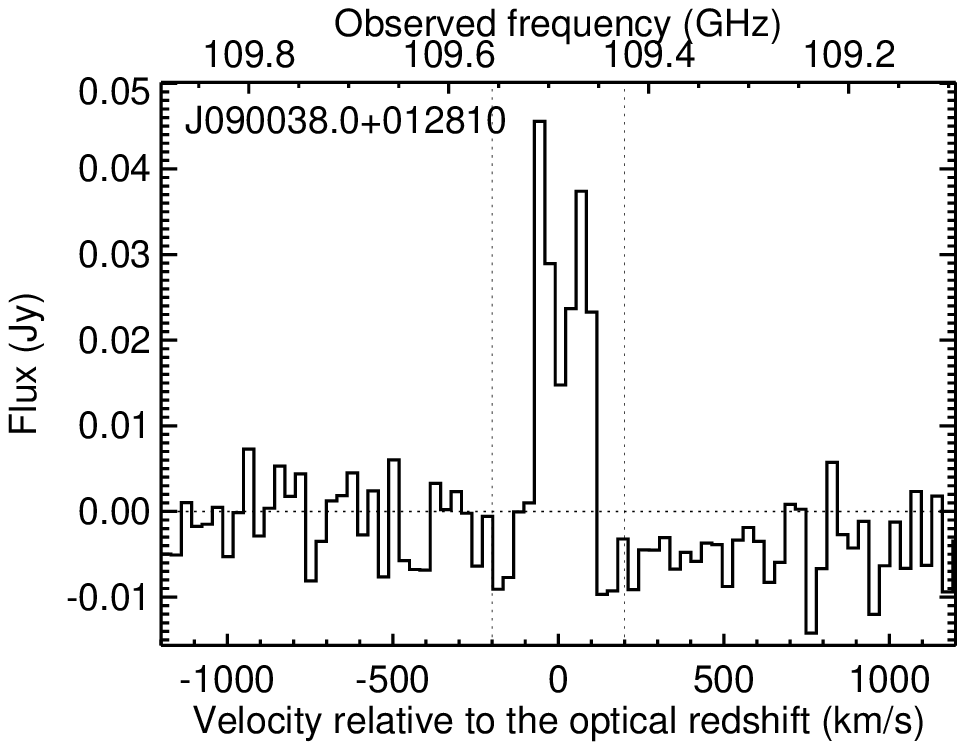}     &  \includegraphics[width=\panelwidth]{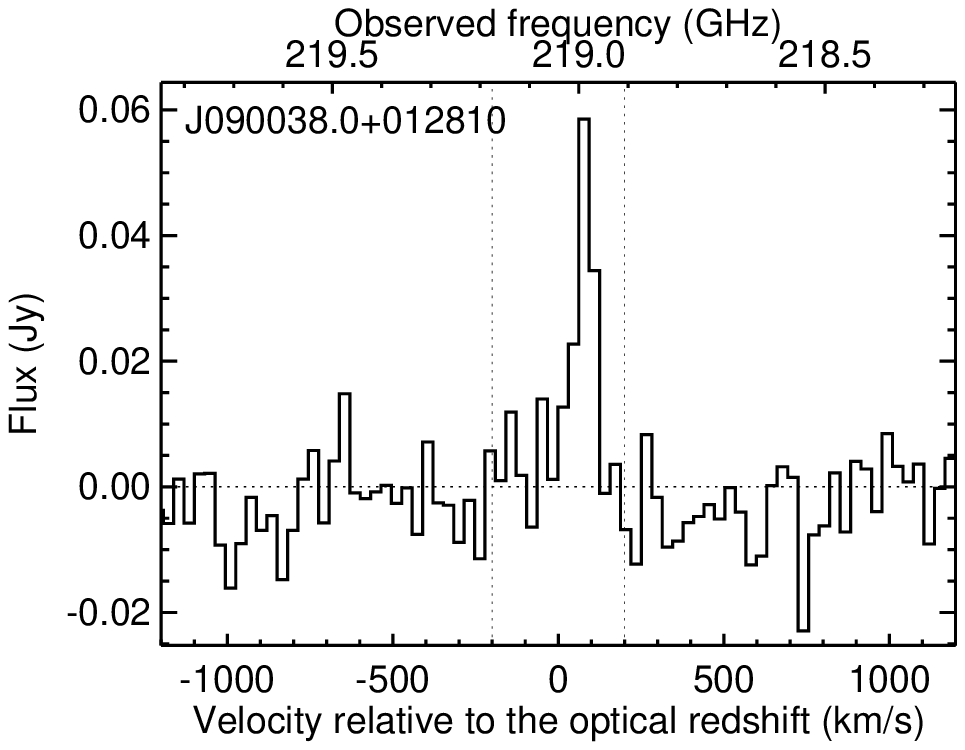} & 
\includegraphics[width=\panelwidth]{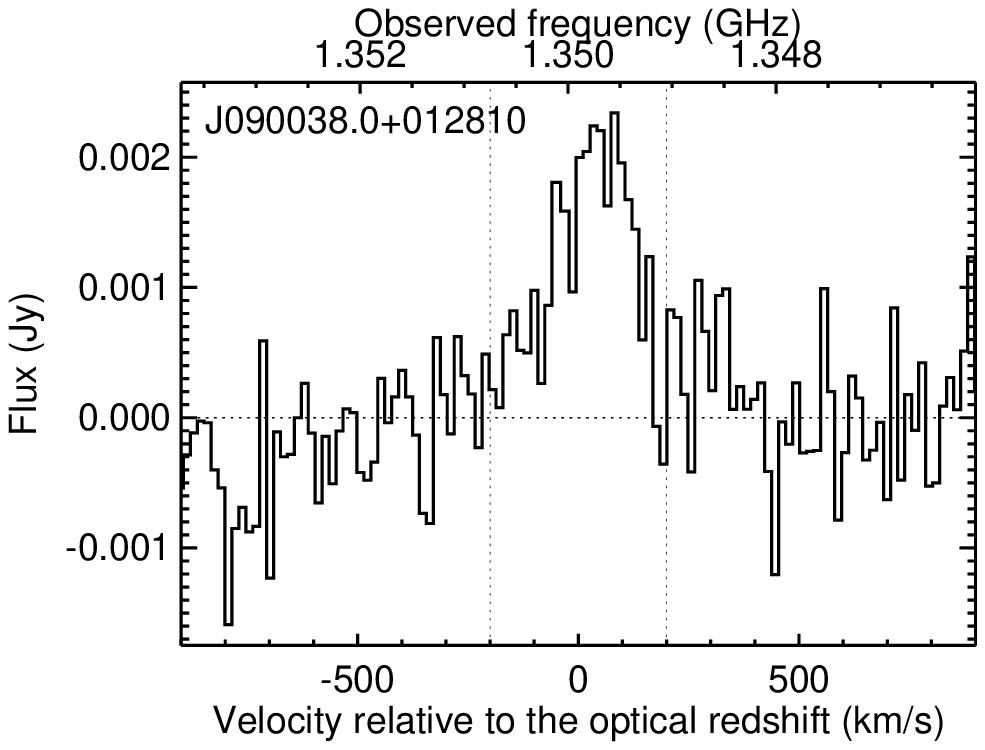} \\
\end{tabular}
\caption{CO(1-0) ({\it left}), CO(2-1) ({\it middle}), and {\hi} ({\it right}) spectra from the IRAM30m and GBT observations. The vertical dotted lines show the velocity range over which the spectra were integrated in order to obtain line fluxes.}
 \label{fig:emirspec}
\end{figure*}

\addtocounter{figure}{-1}

\begin{figure*}
\begin{tabular}{ccc}
\includegraphics[width=\panelwidth]{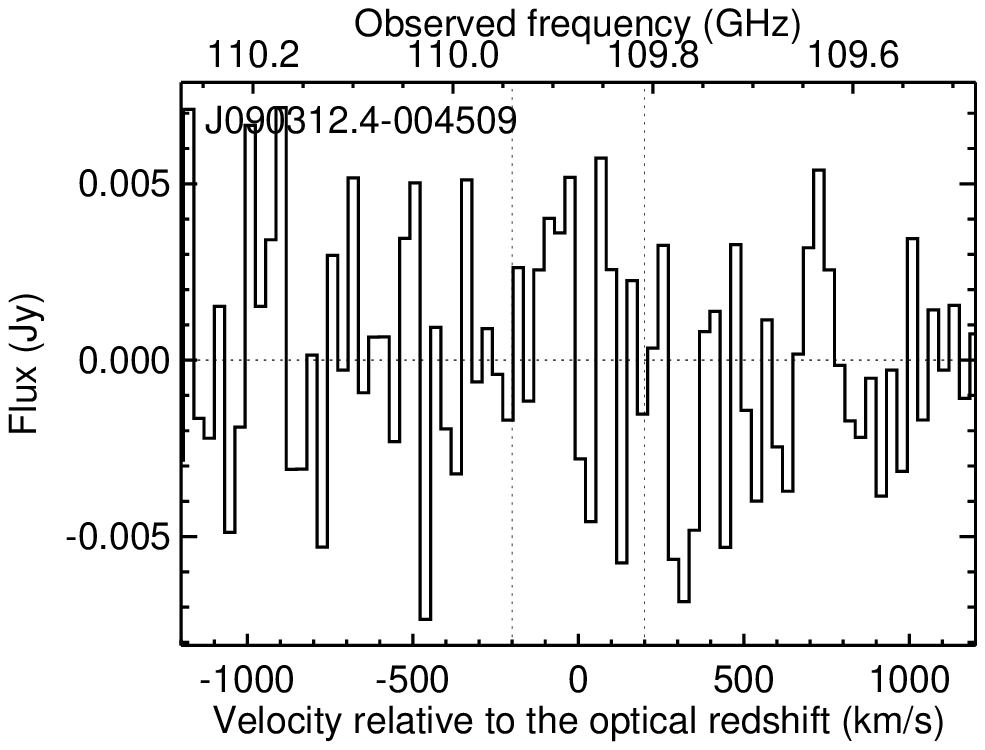}     &  \includegraphics[width=\panelwidth]{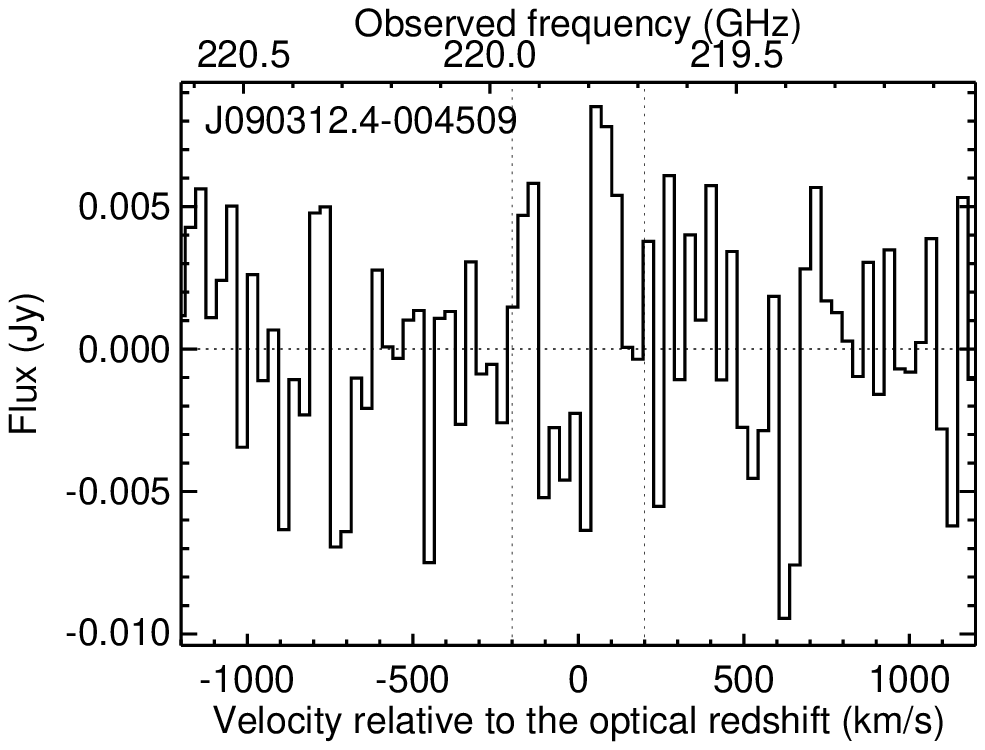} & 
\includegraphics[width=\panelwidth]{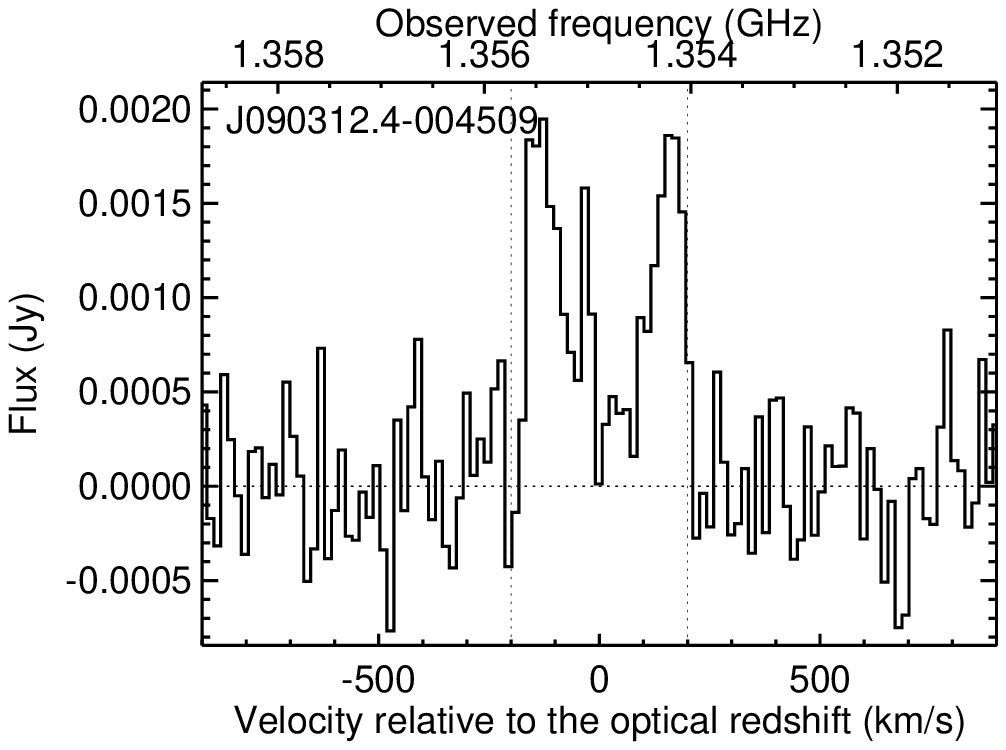} \\
\includegraphics[width=\panelwidth]{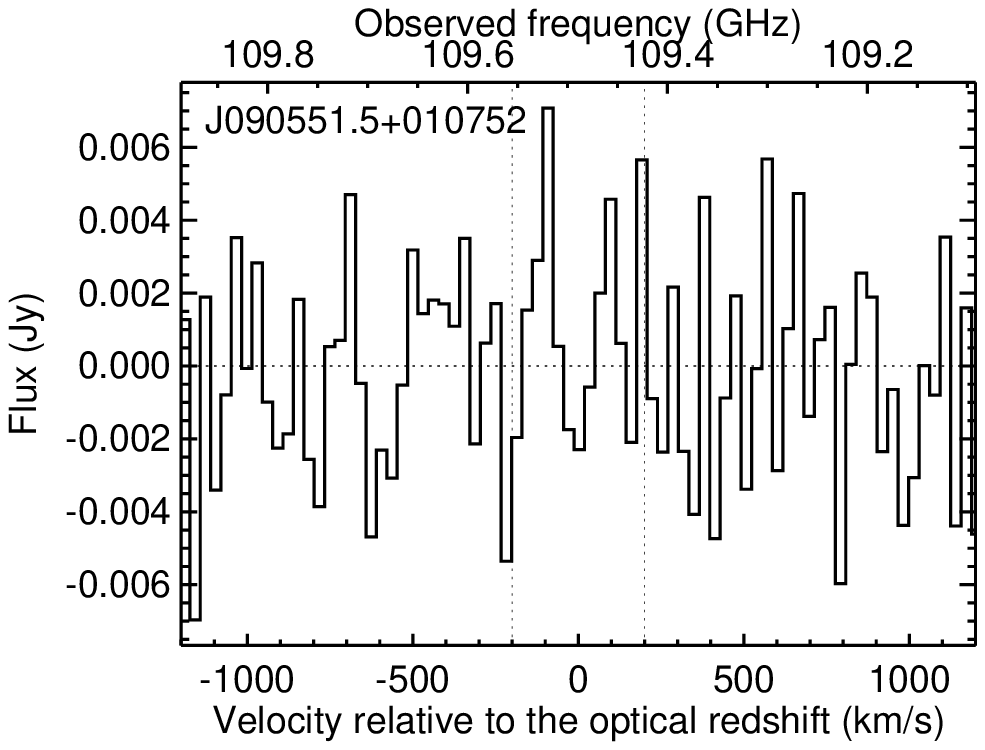}     &  \includegraphics[width=\panelwidth]{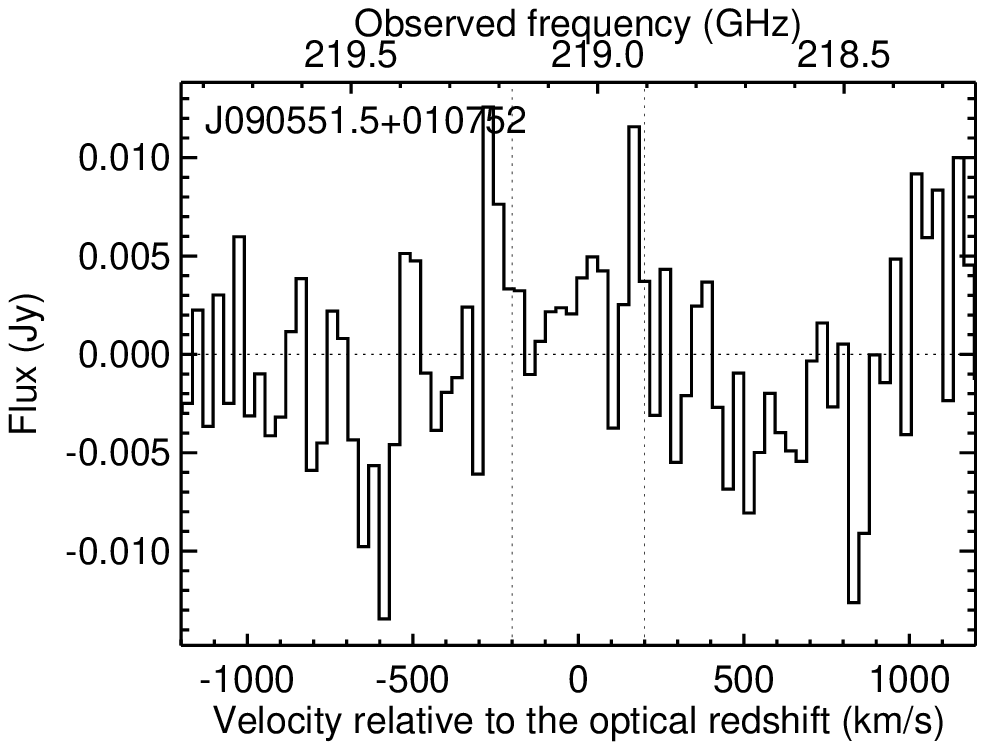} & 
\includegraphics[width=\panelwidth]{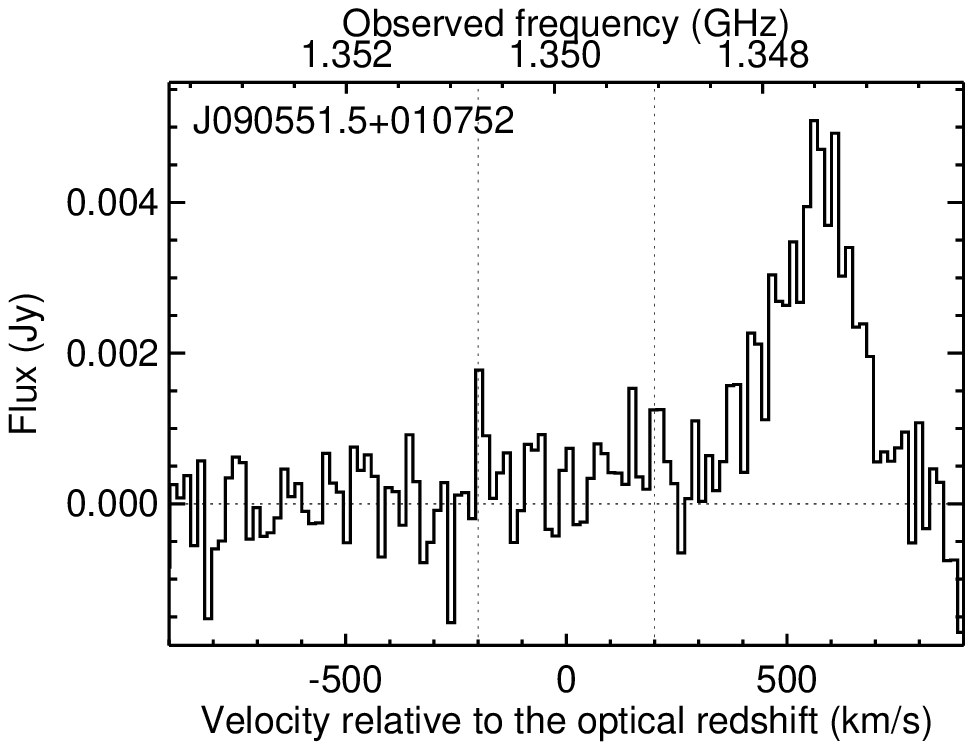} \\
\includegraphics[width=\panelwidth]{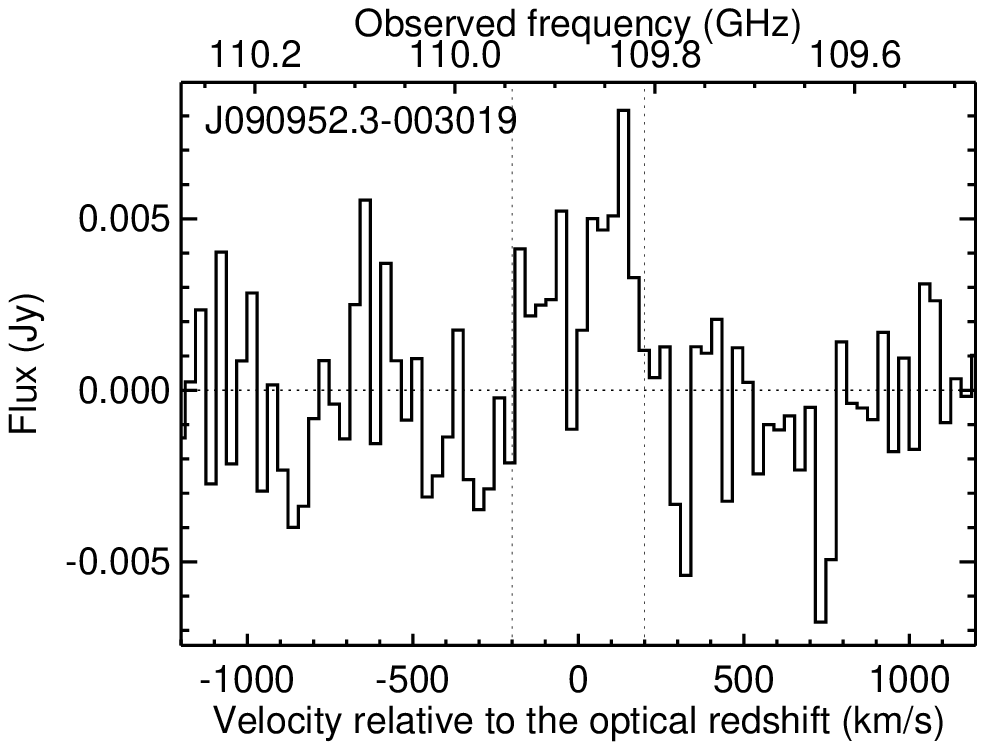}     &  \includegraphics[width=\panelwidth]{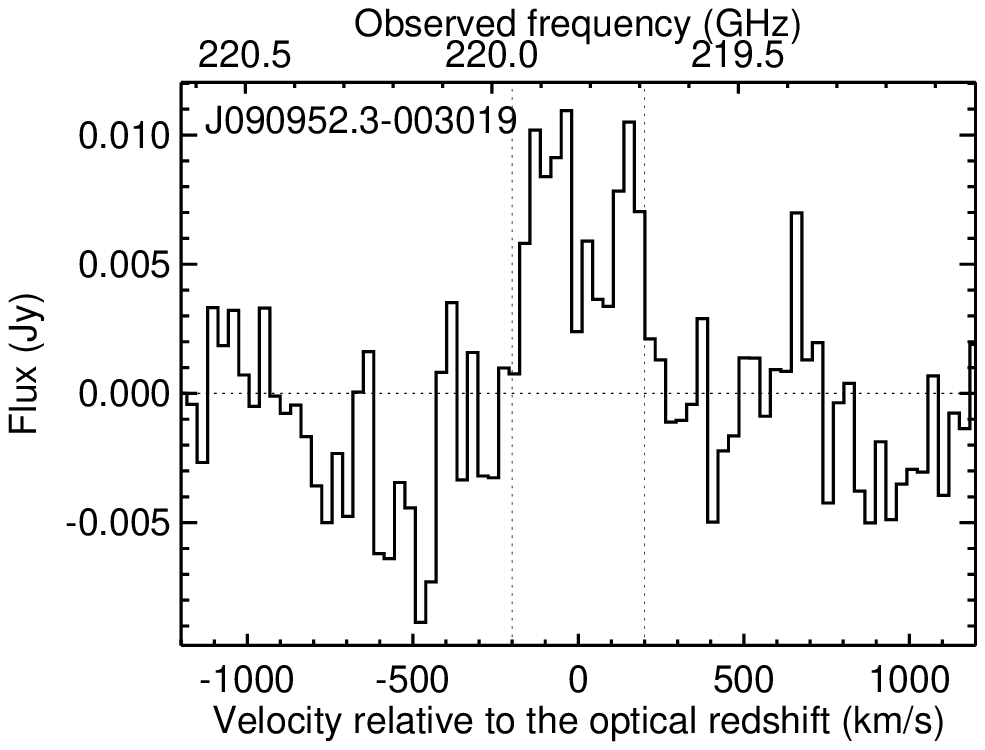}  \\
\includegraphics[width=\panelwidth]{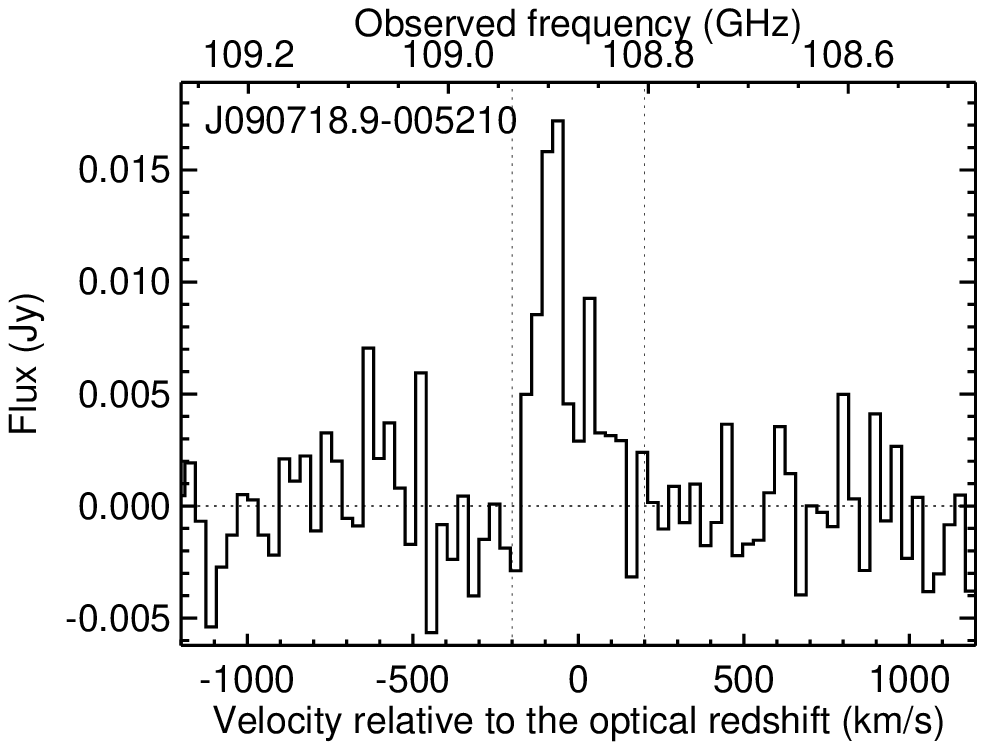}     &  \includegraphics[width=\panelwidth]{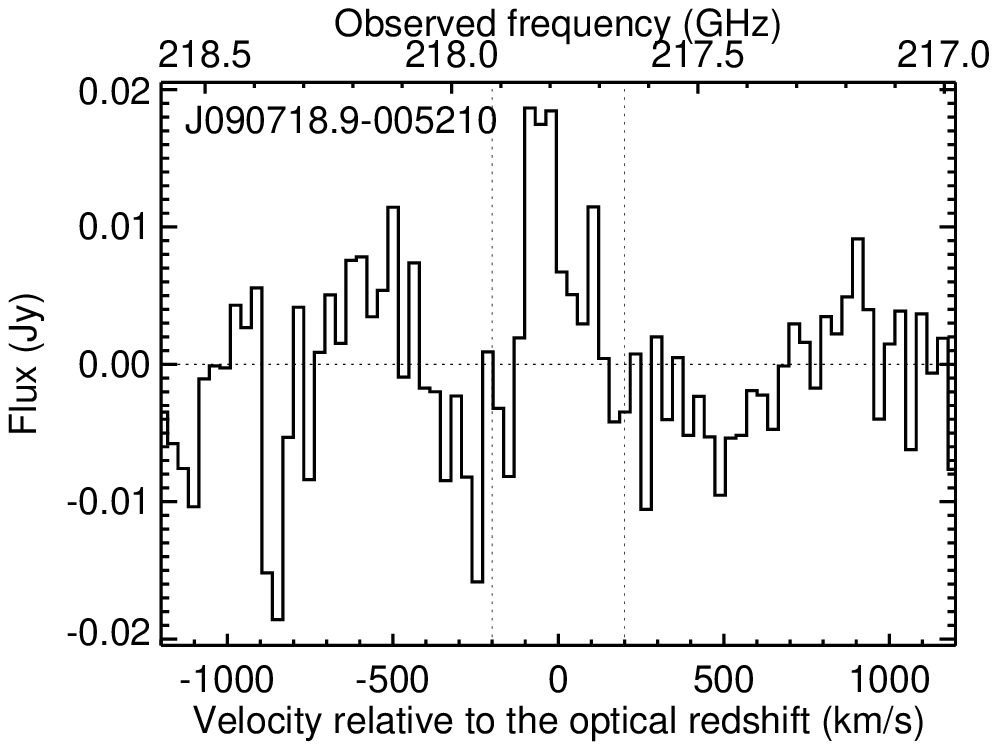} \\
\includegraphics[width=\panelwidth]{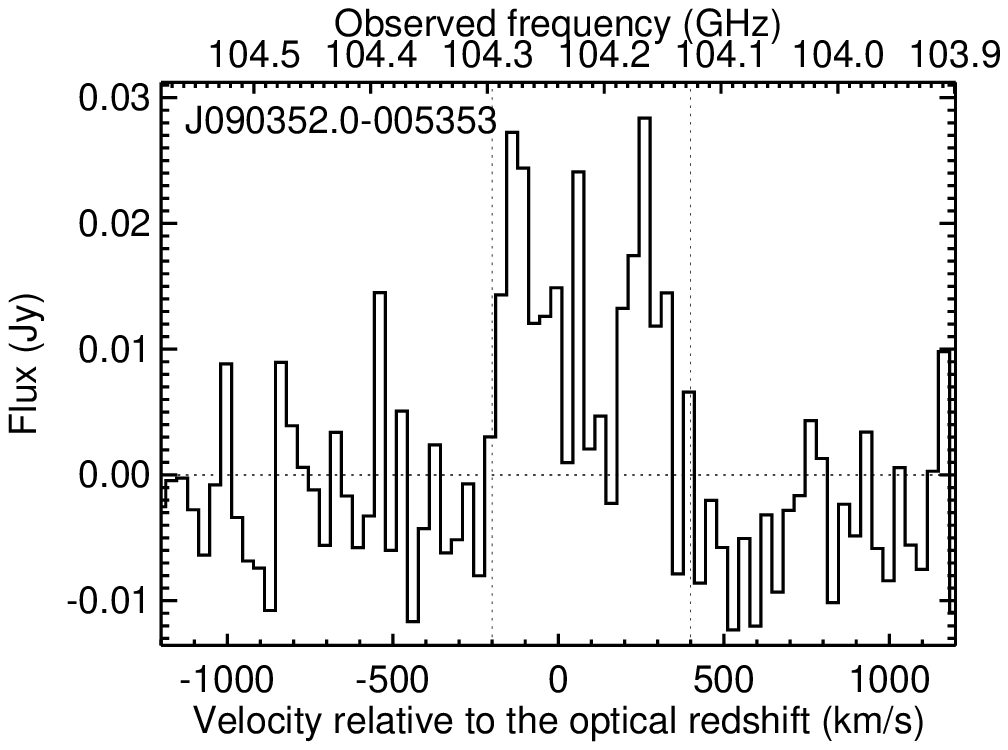}      \\
\includegraphics[width=\panelwidth]{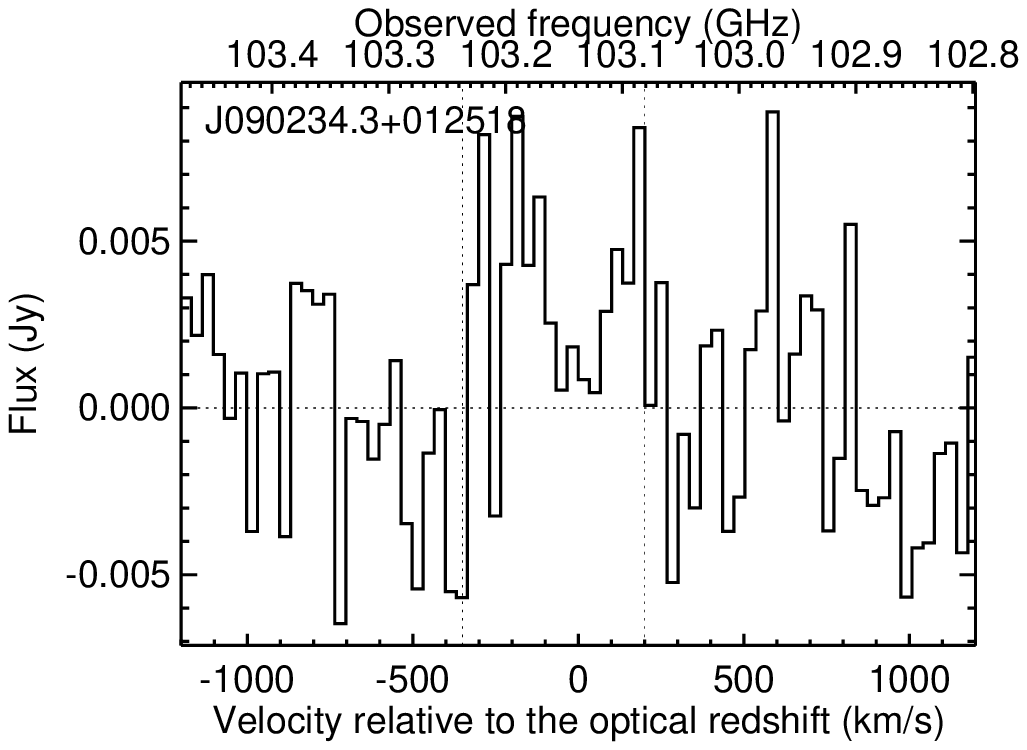} \\
\end{tabular}
\caption{Continued.}
\end{figure*}

\addtocounter{figure}{-1}

\begin{figure*}
\begin{tabular}{ccc}
\includegraphics[width=\panelwidth]{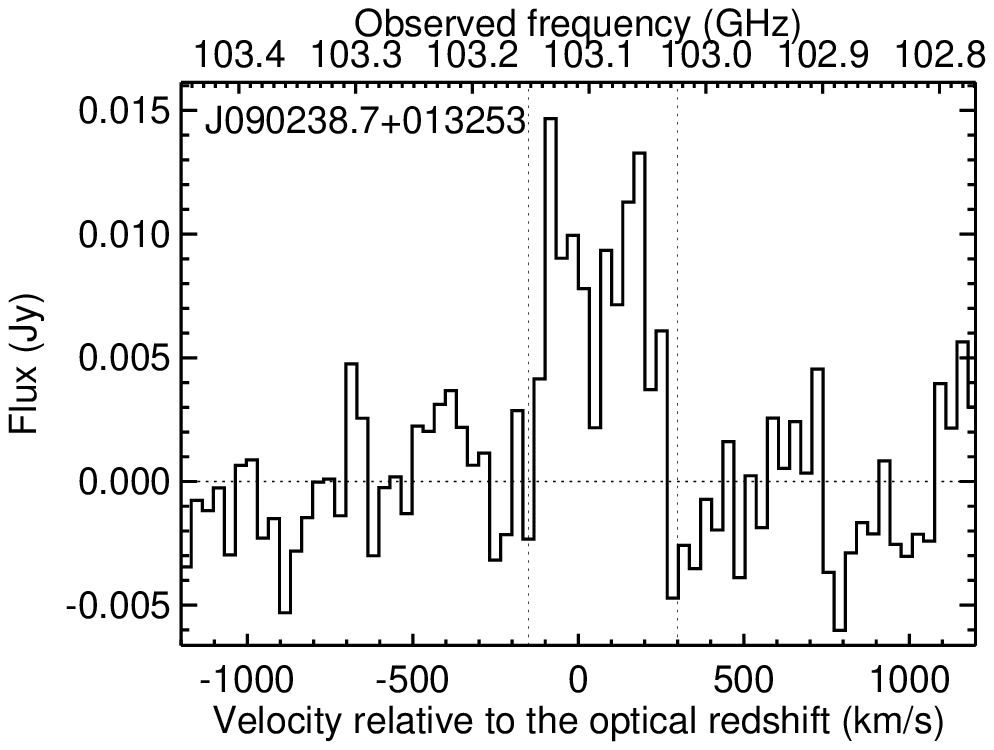} \\
& \hspace{\panelwidth} & \includegraphics[width=\panelwidth]{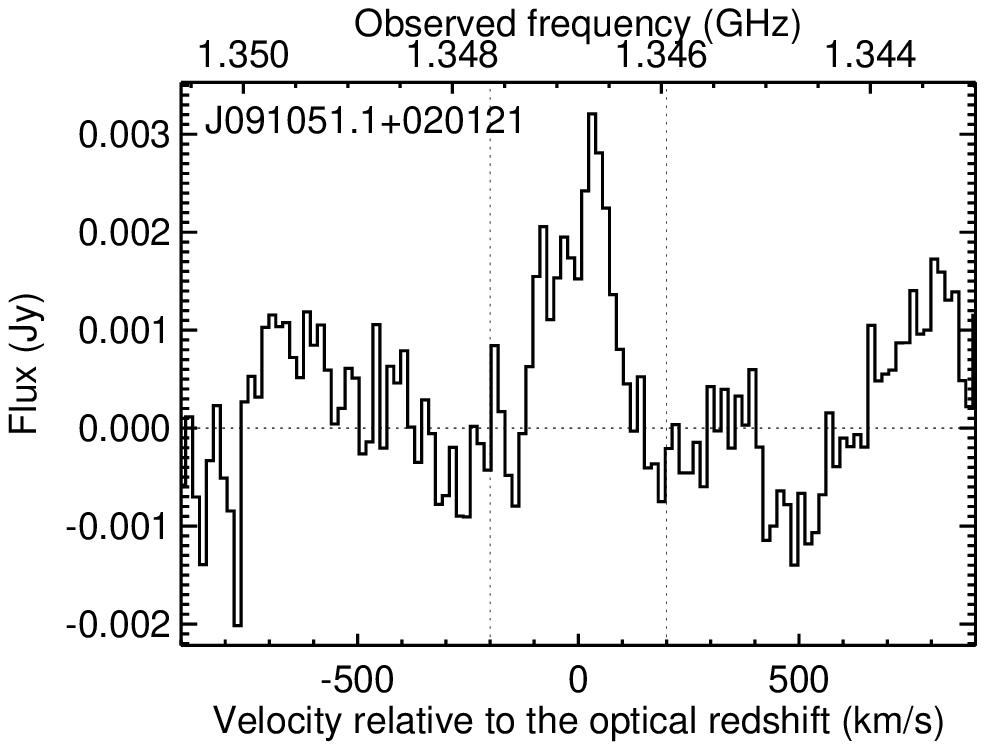} \\
\end{tabular}
\caption{Continued.}
\end{figure*}

\begin{figure*}
\includegraphics[width=0.8\textwidth]{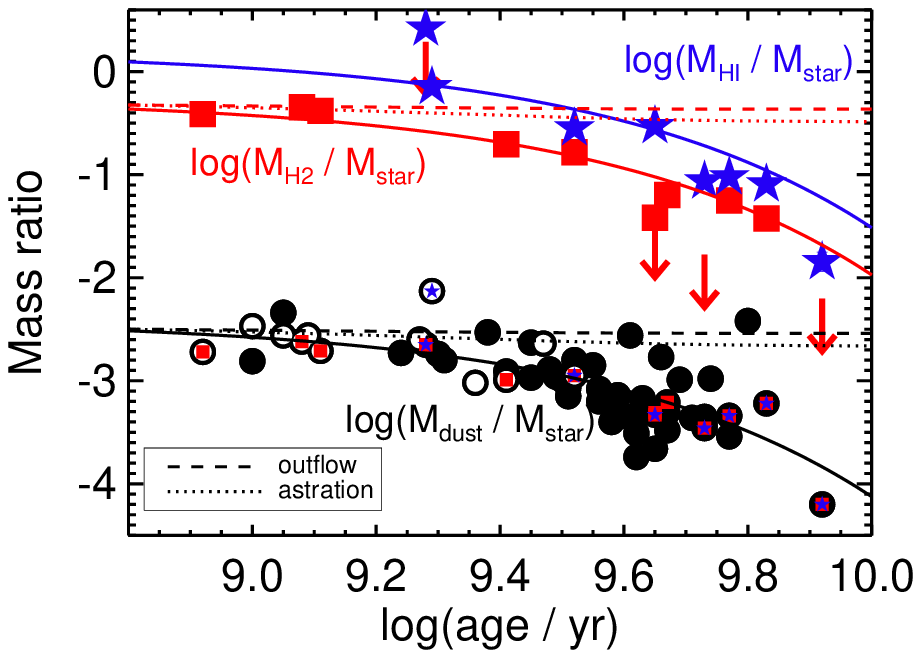}
 \caption{The same as Fig.~\ref{fig:mdms_age}, but showing the effect of outflows and astration. The {\it dotted and dashed lines} denote the molecular gas and dust mass evolution assuming that outflows (with a rate as measured in Fig.~\ref{fig:out_age}) and astration (with the SFR as measured in Fig.~1 of \citetalias{michalowski19etg}) are the only cause of the gas mass change, respectively. This shows that these two effects are too weak to explain the data.
}
 \label{fig:mdms_age_outf_astra}
\end{figure*}

\begin{figure*}
\begin{tabular}{cc}
\includegraphics[width=0.5\textwidth]{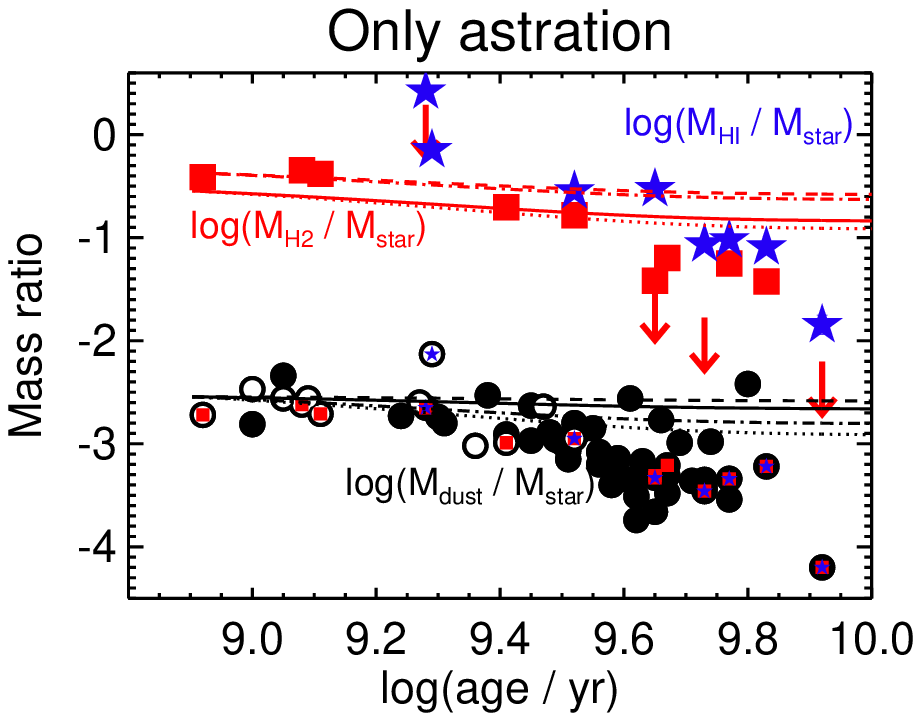} & 
\includegraphics[width=0.5\textwidth]{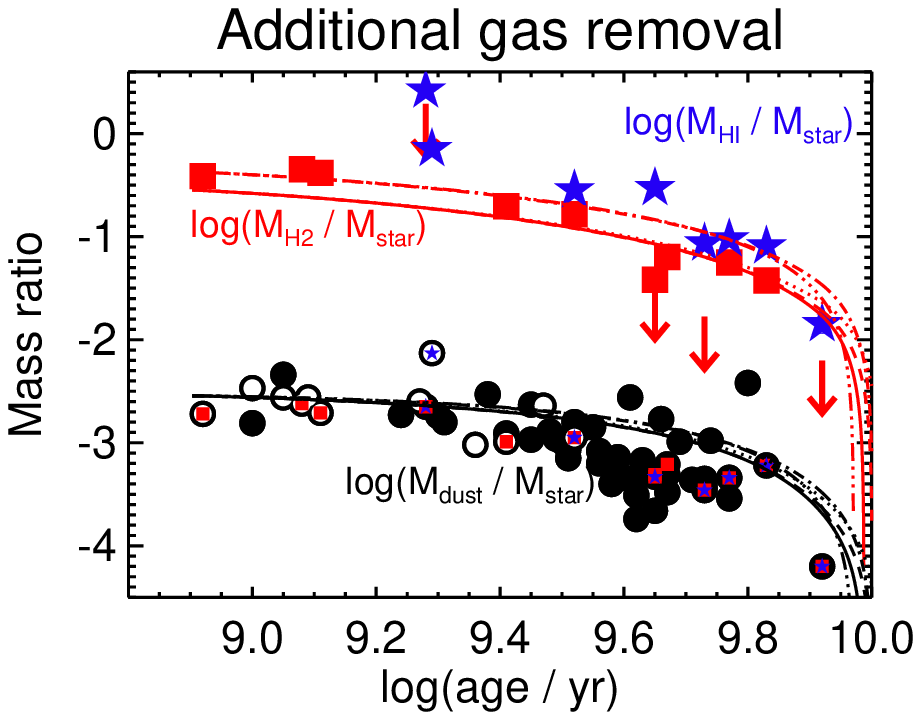} \\
\end{tabular}
 \caption{The same as Fig.~\ref{fig:mdms_age}, but showing the results of the models of \citet{gall11} presented in Section~\ref{sec:mod}. The {\it left} panel show the models with astration (with the SFR as measured in Fig.~1 of \citetalias{michalowski19etg}) as the only mechanism of ISM removal. In the {\it right} panel models with additional cold gas removal (gas heating or outflows) are shown. The model parameters are shown in Table~\ref{tab:demonpar}. This shows that some additional gas removal mechanism is necessary to explain the data.
}
 \label{fig:mdms_age_model}
\end{figure*}

\begin{figure*}
\begin{tabular}{cc}
\includegraphics[height=0.5\textwidth,angle=-90]{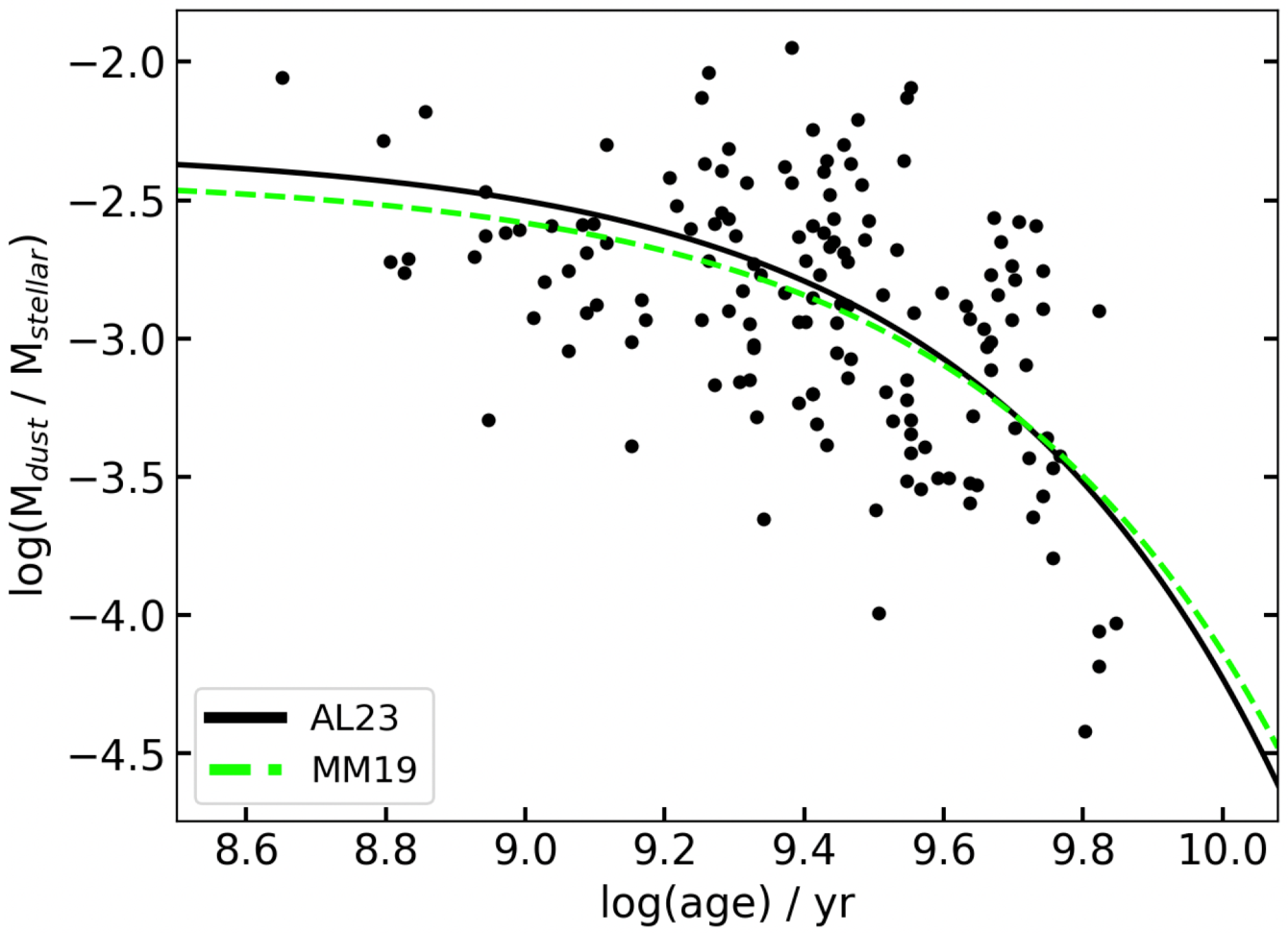} & 
\includegraphics[height=0.5\textwidth,angle=-90]{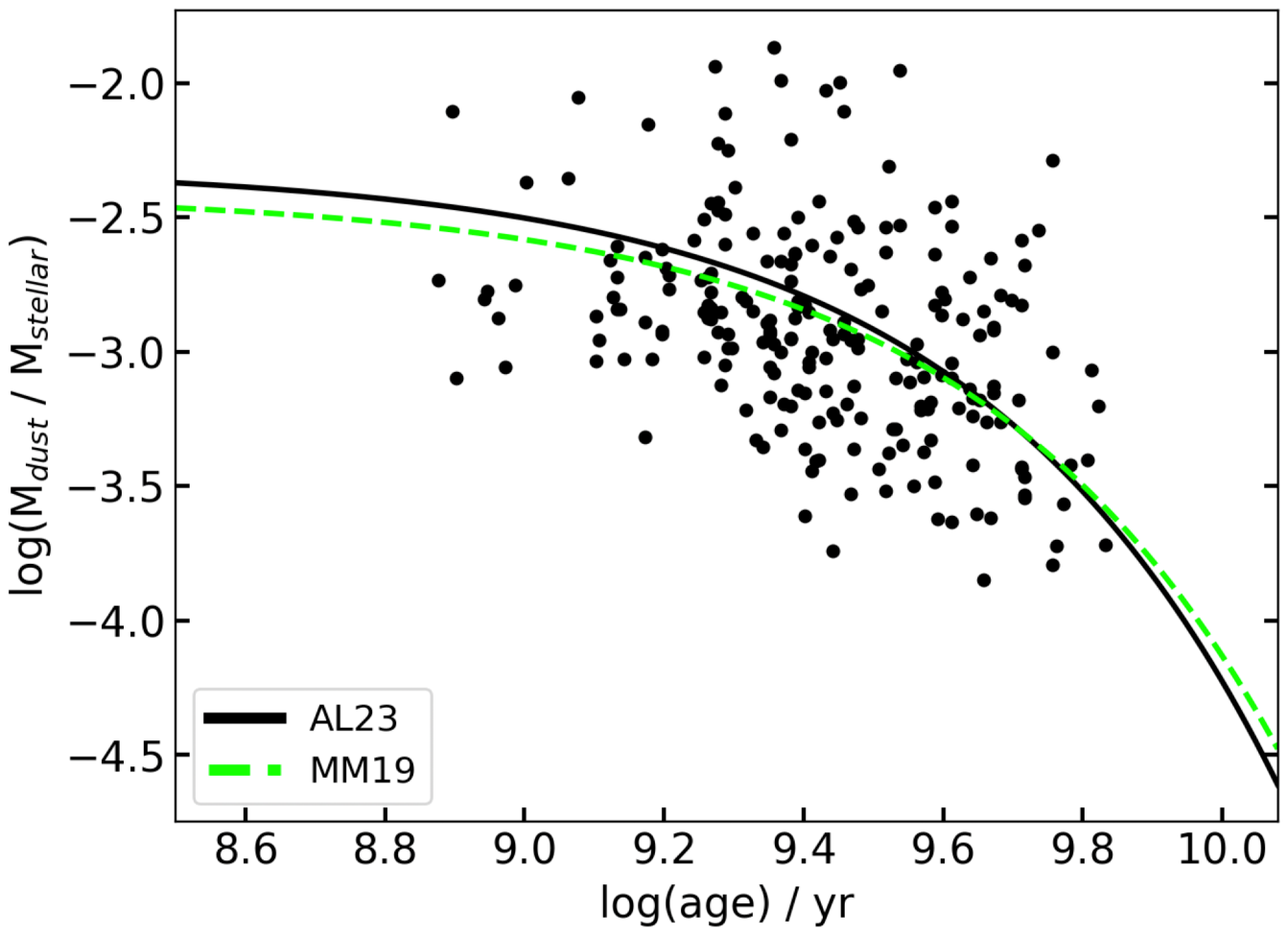} \\
\end{tabular}
 \caption{The same as Fig.~\ref{fig:mdms_age}, but for the larger sample of \citet{lesniewska23} only for a narrow range of stellar masses: $\log(\mstar/\msun)=10.5$--$10.6$ ({\it left}) and $10.6$--$10.7$ ({\it right}). The {\it solid black line} shows the fit to the full sample of \citet{lesniewska23} and the {\it dashed green line} shows the fit to the sample in \citet{michalowski19etg}. This shows that the $\mdust/\mstar$-age anti-correlation is not driven by the $\mstar$-age correlation, because it would disappear for narrow ranges of $\mstar$. 
}
 \label{fig:mdms_age_narrowms}
\end{figure*}

\input{table_model_parameters}


\end{document}

%% file: COflux.tex
\begin{table*}
\caption{CO fluxes and luminosities from the IRAM30m/EMIR observations.}
\label{tab:emirres}
\begin{center}
\scriptsize
\begin{tabular}{lcccc|cccc}
\hline\hline
Galaxy & \multicolumn{4}{c|}{CO(1-0)} & \multicolumn{4}{c}{CO(2-1)} \\
	     & FWHM & $F_{\rm int}$ & $\lp$ & $\mhtwo$ 
	     & FWHM & $F_{\rm int}$ & $\lp$ & $\mhtwo$ \\
       &(\kms)& (Jy \kms) & ($10^{8}\mbox{K km s}^{-1} \mbox{ pc}^2$) & ($10^8\msun$) 
       &(\kms)& (Jy \kms) & ($10^{8}\mbox{K km s}^{-1} \mbox{ pc}^2$) & ($10^8\msun$) \\
\hline
J091205.8+002656 & $158 \pm 20$ &$10.46 \pm 1.38$ &$14.32 \pm 1.89$ &$71.6 \pm 9.5$  & $226 \pm 12$ &$19.41 \pm 3.37$ &$6.64 \pm 1.15$ &$66.4 \pm 11.5$  \\
J091448.7-003533 & $268 \pm 22$ &$6.94 \pm 1.32$ &$9.31 \pm 1.76$ &$46.5 \pm 8.8$  & $194 \pm 71$ &$5.59 \pm 2.70$ &$1.87 \pm 0.91$ &$18.7 \pm 9.1$  \\
J085828.5+003814 & $235 \pm 16$ &$6.17 \pm 1.15$ &$7.79 \pm 1.45$ &$39.0 \pm 7.2$  &  $\cdots$ &$9.48 \pm 3.95$ &$<5.49$ &$<54.9$  \\
J085946.7-000020 &  $\cdots$ &$0.60 \pm 0.74$ &$<2.67$ &$<13.3$  &  $\cdots$ &$0.87 \pm 1.12$ &$<1.00$ &$<10.0$  \\
J085915.7+002329 &  $\cdots$ &$0.01 \pm 0.52$ &$<0.06$ &$<0.3$  &  $\cdots$ & $\cdots$ & $\cdots$ & $\cdots$  \\
J090038.0+012810 & $235 \pm 10$ &$4.33 \pm 0.45$ &$5.44 \pm 0.57$ &$27.2 \pm 2.8$  & $225 \pm 28$ &$4.84 \pm 0.71$ &$1.52 \pm 0.22$ &$15.2 \pm 2.2$  \\
J090312.4-004509 &  $\cdots$ &$0.40 \pm 0.39$ &$<1.27$ &$<6.4$  &  $\cdots$ &$0.39 \pm 0.42$ &$<0.33$ &$<3.3$  \\
J090551.5+010752 &  $\cdots$ &$0.45 \pm 0.34$ &$<1.42$ &$<7.1$  &  $\cdots$ &$1.12 \pm 0.63$ &$<0.75$ &$<7.5$  \\
J090952.3-003019 & $237 \pm 56$ &$1.36 \pm 0.27$ &$1.46 \pm 0.29$ &$7.3 \pm 1.4$  & $235 \pm 22$ &$2.68 \pm 0.39$ &$0.72 \pm 0.11$ &$7.2 \pm 1.1$  \\
J090718.9-005210 & $242 \pm 18$ &$2.19 \pm 0.32$ &$3.46 \pm 0.51$ &$17.3 \pm 2.6$  & $168 \pm 19$ &$2.12 \pm 0.69$ &$0.84 \pm 0.27$ &$8.4 \pm 2.7$  \\
J090352.0-005353 & $427 \pm 60$ &$7.26 \pm 0.81$ &$36.10 \pm 4.04$ &$180.5 \pm 20.2$  &  $\cdots$ & $\cdots$ & $\cdots$ & $\cdots$  \\
J090234.3+012518 & $302 \pm 10$ &$1.83 \pm 0.52$ &$11.59 \pm 3.32$ &$58.0 \pm 16.6$  &  $\cdots$ & $\cdots$ & $\cdots$ & $\cdots$  \\
J090238.7+013253 & $254 \pm 16$ &$3.13 \pm 0.33$ &$19.97 \pm 2.09$ &$99.9 \pm 10.5$  &  $\cdots$ & $\cdots$ & $\cdots$ & $\cdots$  \\
\hline
\end{tabular}
\tablefoot{The columns show the Gaussian FWHM of the line, the integrated line flux within the dotted lines on Fig.~\ref{fig:emirspec}, the line luminosity and the molecular gas mass assuming $\alpha_{\rm CO}=5\,M_\odot\, (\mbox{K km s}^{-1} \mbox{ pc}^2)^{-1}$. The left set is for CO(1-0) and the right set is CO(2-1) for galaxies for which simultaneous setting for this line was possible.}
\end{center}
\end{table*}

%% file: HIflux.tex
\begin{table*}
\caption{{\hi} fluxes and luminosities from the GBT/VEGAS observations.}
\label{tab:gbtres}
\begin{center}
\begin{tabular}{lcccc}
\hline\hline
Galaxy & FWHM & $F_{\rm int}$ & $\lp$ & $\mhi$ \\
       &(\kms)& (Jy \kms) & ($10^{10}\mbox{K km s}^{-1} \mbox{ pc}^2$) & ($10^8\msun$) \\
\hline
J085828.5+003814 & $243 \pm 15$ &$0.674 \pm 0.062$ &$56.12 \pm 5.16$ &$86.52 \pm 7.95$  \\
J085915.7+002329 & $188 \pm 2$ &$1.960 \pm 0.038$ &$7.44 \pm 0.14$ &$11.02 \pm 0.21$  \\
J085946.7-000020 & $464 \pm 9$ &$1.377 \pm 0.044$ &$116.47 \pm 3.69$ &$179.64 \pm 5.69$  \\
J090038.0+012810 & $233 \pm 15$ &$0.460 \pm 0.042$ &$38.08 \pm 3.44$ &$58.71 \pm 5.30$  \\
J090312.4-004509 & $322 \pm 13$ &$0.390 \pm 0.030$ &$27.63 \pm 2.11$ &$42.43 \pm 3.25$  \\
J090551.5+010752 &  $\cdots$ &$0.169 \pm 0.041$ &$<20.83$ &$<32.11$  \\
J091051.1+020121 & $173 \pm 11$ &$0.377 \pm 0.054$ &$34.20 \pm 4.87$ &$52.84 \pm 7.52$  \\
J091448.7-003533 & $398 \pm 26$ &$0.587 \pm 0.048$ &$51.85 \pm 4.28$ &$80.05 \pm 6.61$  \\
\hline
\end{tabular}
\tablefoot{The columns show the Gaussian FWHM of the line, the integrated line flux within the dotted lines on Fig.~\ref{fig:emirspec}, the line luminosity and the atomic gas mass.}
\end{center}
\end{table*}

%% file: table_model_parameters.tex
\begin{table*}
\caption{Chemical evolution model main parameters.}
\label{tab:demonpar}
\begin{center}
\begin{tabular}{lllll}
\hline\hline
 M$_{ISM}$	& SN$^a$    &  AGB$^a$ 	& M$_{out}$ & t$_{out}$ \\ 
 $10^{10} \msun$	& recycling & recycling	& $10^{10} \msun$  & $10^{10}$ year  \\ 
\hline

 \multicolumn{5}{c}{Only astration}\\
 2.0  & $\surd$ & no      & $\cdots$  & $\cdots$  \\ 
 2.0  & no      & no      & $\cdots$  & $\cdots$  \\    	
 3.0  & $\surd$ & no      & $\cdots$  & $\cdots$  \\ 	
3.0  & no      & no      & $\cdots$  & $\cdots$  \\	
 \hline
 \multicolumn{5}{c}{With additional gas removal}\\
2.0  & $\surd$ &  $\surd$ & 2.9 & 1.0 \\
2.0  & $\surd$ &  $\surd$ & 2.8 & 1.0 \\
2.0  & $\surd$ &  $\surd$ & 2.7 & 0.9 \\
3.0  & $\surd$ &  $\surd$ & 4.4 & 0.9 \\
3.0  & $\surd$ &  $\surd$ & 5.4 & 1.2 \\
\hline 
\end{tabular}
\tablefoot{$^a$ Refers to whether the mass recycled from dying stars is considered to be instantaneously available for star formation or not.}
\end{center}
\end{table*}